\RequirePackage{fix-cm}
\documentclass[smallextended]{svjour3}       
\smartqed  
\usepackage[utf8]{inputenc}
\usepackage[english]{babel}

\usepackage{amsmath}
\usepackage{amsthm}
\usepackage{mathtools}
\usepackage{amsfonts}
\usepackage{bbold}
\usepackage{amssymb}
\usepackage{graphicx}
\usepackage{natbib}
\usepackage{float}
\usepackage[citecolor=blue,]{hyperref}
\usepackage{lscape}
\begin{document}

\title{Analytic expressions for the Cumulative Distribution Function of the Composed Error Term in Stochastic Frontier Analysis with Truncated Normal and Exponential Inefficiencies
}
\subtitle{Four Representation Theorems}


\author{Rouven Schmidt         \and
        Thomas Kneib 
}


\institute{Rouven Schmidt \at
              Georg-August University G\"ottingen \\
              Tel.: +49 (0) 551 - 39 25431\\ 
              \email{rouven.schmidt@uni-goettingen.de}           
           \and
           Thomas Kneib \at
              Georg-August University G\"ottingen \\
              Tel.: +49 (0) 551 / 39 25678\\ 
              \email{tkneib@uni-goettingen.de}           
}

\date{Received: date / Accepted: date}

\maketitle

\begin{abstract}
In the stochastic frontier model the composed error term consists of the measurement error and the inefficiency term. A general assumption is that the inefficiency term follows a truncated normal or exponential distribution. In a wide variety of models evaluating the cumulative distribution function of the composed error term is required. This work introduces and proves four representation theorems for these distributions - two for each distributional assumptions. These representation can be utilized for a fast and accurate evaluation.
\end{abstract}

\newpage

\section{Introduction}
\label{intro}
In the stochastic frontier model, the composed error term consists of the measurement error $v \sim N(0,\sigma_v^2)$ and the inefficiency term $u$.
The inefficiency $u$ is assumed to be greater or equal to zero and random. Thus a random variable with a positive support is facilitated to model $u$.  If one assumes independence of $v$ and $u$, the composed error for production inefficiency is defined as $\epsilon= v - u$. 
The cumulative distribution function (cdf) of $\epsilon$ is specified as:
\[
F_\epsilon(\kappa)=\int_{-\infty}^{\kappa} f_\epsilon(t) dt.
\]
The cost inefficiency composed error term is defined as $\epsilon^*= v + u$. Thus the cdf of $\epsilon^*$ can be written as: $F_{\epsilon^*}(\kappa)=1-F_{\epsilon}(-\kappa)$ \footnote{The proof is analogous to the one of Theorem \ref{theoremexpcdfemg}}. \cite{meesters2014note} notes that common assumptions for the distribution of the inefficiency terms are the truncated (below zero) normal , exponential and half-normal distribution. If $u$ is assumed to follow a truncated (below zero) normal distribution, i.e. $u \sim TN(\mu,\sigma_u,0,\infty)$ 
,then the probability density function(pdf) of $\epsilon$ is derived by \cite{kumbhakar2015practitioner} as:
\begin{equation} \label{tnpdf}
f_\epsilon(\epsilon) = \frac{1}{\sqrt{\sigma_v^2 + \sigma_u^2} \Phi (\frac{\mu}{\sigma_u})} \phi \left( \frac{\epsilon+\mu}{\sqrt{\sigma_v^2 + \sigma_u^2}}\right) \Phi \left( \frac{  \mu \sigma_v^2 - \epsilon \sigma_u^2 }{\sqrt{\sigma_v^2 + \sigma_u^2} \sigma_v \sigma_u} \right)
\end{equation}
with $\mu \in \mathbb{R}$ , $\sigma >0$. Further, $\phi(\cdot)$ and $\Phi(\cdot)$ are the pdf and cdf of the standard normal distribution respectively. Setting $\mu=0$ in the truncated normal distribution yields the half-normal distribution, thus the truncated normal distribution is a generalization of the half-normal distribution.  Consequently, the truncated normal distribution is more flexible, with the trade-off of having one additional parameter. It was first introduced by \cite{stevenson1980likelihood}.
\newline 
Alternatively assume that the random variable $u$ follows an exponential distribution, i.e. $u \sim Exp(\lambda)$ where $\lambda >0$, 
then the pdf of $\epsilon$ is given by:
\begin{equation} \label{exppdf}
f_\epsilon(\epsilon)=\lambda \exp \{ \lambda \epsilon + \frac{\sigma_v^2 \lambda^2 }{2}  \} \Phi \left( -\frac{\epsilon}{\sigma_v}-\lambda \sigma_v \right)
\end{equation}
see \cite{kumbhakar2015practitioner}. The mode of the distribution is at $0$, thus implying  the mode of producers to be efficient. This approach to inefficiency modeling was first introduced by \cite{meeusen1977efficiency}. \\
Recently more and more models are developed, which do not only require the pdf but also the cdf of the random error to estimate the model parameters. 
Examples are  \cite{genius2012measuring}, \cite{lai2013maximum}, \cite{tsay2013simple}, \cite{amsler2014using}, \cite{tran2015endogeneity} and \cite{sriboonchitta2017double}. The recent paper by \cite{amsler2019evaluating} introduced a representation of the cdf of the composed error term assuming $u$ follows the half-normal distribution. Before that, one had to rely on numerical integration methods to evaluate $F_\epsilon(\cdot)$ \footnote{\cite{lai2013maximum} introduced a numerical approximation, which breaks down for some parameter combinations. It also contained a typo so that for $\lim_{\kappa \to -\infty} F_\epsilon(\kappa) \neq 0$. The correction of which is supplied by Lai and Huang if requested.}. Utilizing analytical representations of integrals is generally more accurate and yields a faster computation which thus allows for the estimation of more complex models. This work introduces the cdfs of $\epsilon$ for inefficiency terms $u$ following a truncated normal or exponential distribution. Section \ref{Truncated Normal Model} introduces two separate representation theorems for the composed error term involving a truncated normal distribution.  In Section \ref{Exponential Model}, two theorems are introduced that allow to analytically represent $F_\epsilon(\cdot)$ if $u$ follows and exponential distribution. 
The proof of all theorems and lemmas are provided. The findings are then validated through simulation in Section \ref{Simulation}.  

\section{Truncated Normal Inefficiency Model} \label{Truncated Normal Model}
In the following section two representations of $F_\epsilon(\cdot)$ with $u \sim TN(\mu,\sigma_u, 0, \infty)$ are introduced. Further, the proofs are provided. Additionally, information on the limiting behavior is given.

\subsection{Representation using Owen's T function }
\begin{theorem}  \label{theoremtncdf}
Let $u \sim TN(\mu,\sigma_u, 0, \infty)$ and $v \sim N(0, \sigma_v)$ be independent, then it holds that the cdf of $\epsilon=v-u$ can be represented as:
\begin{align*}
F_\epsilon(\kappa) &=\frac{1}{\Phi (\frac{\mu}{\sigma_u})} \Bigg[ \frac{1}{2}\Phi(\varphi(\kappa))+\frac{1}{2}\Phi(\frac{a}{\sqrt{1+b^2}}) -\frac{1}{2} \mathbb{1}_{(-\infty,0)}(\frac{a}{\varphi(\kappa) \sqrt{1+b^2}})\\
&-T \left( \varphi(\kappa),\frac{a+b\varphi(\kappa)}{\varphi(\kappa)} \right)-T \left( \frac{a}{\sqrt{1+b^2}}, \frac{ab+\varphi(\kappa)(1+b^2)}{a} \right) \Bigg] \\
\end{align*} 
where $a=\frac{\mu \sqrt{\sigma_v^2 + \sigma_u^2} }{\sigma_v \sigma_u}$, $b=- \frac{ \sigma_u }{\sigma_v }$ and $\varphi(\kappa)= \frac{t+\mu}{\sqrt{\sigma_v^2 + \sigma_u^2}}$.  Further, $T(\cdot)$ denotes Owen's T function defined as:
\[
T(h,g)=\frac{1}{2\pi}\int_0^g \frac{\exp \{-\frac{1}{2}h^2(1+t^2) \}}{1+t^2} dt
\]
with $h,g \in \mathbb{R}$ in \cite{owen1956tables}.
\end{theorem}

\newpage

Theorem 1 is a direct consequence of the following Lemma:
\begin{lemma} \label{lemmatncdf}
Let $u \sim TN(\mu,\sigma_u,0 , \infty)$ and $v \sim N(0, \sigma_v)$ be independent, then it holds that the cdf of $\epsilon=v-u$ can be represented as:
\begin{align*}
\int_{-\infty}^{\kappa} f_\epsilon(t) dt &=\frac{1}{\Phi (\frac{\mu}{\sigma_u})} \int_{-\infty}^{\varphi(\kappa)}    \phi \left( y \right) \Phi \left( a+by\right)  dy\\
\end{align*} 
where $a=\frac{\mu \sqrt{\sigma_v^2 + \sigma_u^2} }{\sigma_v \sigma_u}$, $b=- \frac{ \sigma_u }{\sigma_v }$ and $\varphi(\kappa)= \frac{t+\mu}{\sqrt{\sigma_v^2 + \sigma_u^2}}$.
\end{lemma}

\subsubsection{Proof of Lemma \ref{lemmatncdf}}
Initially Lemma \ref{lemmatncdf} is proven and then it is shown how Theorem \ref{theoremtncdf} follows. \newline
\begin{proof}
Given the cdf as constructed through the integral of Equation \ref{tnpdf}, the expression may be simplified by substition:

\[
y= \varphi(t):= \frac{t+\mu}{\sqrt{\sigma_v^2 + \sigma_u^2}} \hspace{0.5cm},
\]
 which can be rearranged as: 
\begin{equation} \label{substitution variable rearranged}
t= y\sqrt{\sigma_v^2 + \sigma_u^2}-\mu \hspace{0.5cm}.
\end{equation}
The derivative of $y$ w.r.t. $t$ is:
\begin{align*}
\frac{d y}{d t}= \frac{1}{\sqrt{\sigma_v^2 + \sigma_u^2}} \Leftrightarrow dt=\sqrt{\sigma_v^2 + \sigma_u^2} dy
\end{align*}

Appropriately transforming the limits of the integral results in:
\begin{align*}
&\varphi(\kappa) = \frac{\kappa+\mu}{\sqrt{\sigma_v^2 + \sigma_u^2}} & \lim_{\kappa \to \infty} \varphi(-\kappa)=-\infty
\end{align*}

and finally introducing $a$ and $b$ for ease of representation:
\begin{align*}
& \frac{\mu \sigma_v^2 - t \sigma_u^2}{\sqrt{\sigma_v^2 + \sigma_u^2} \sigma_v \sigma_u} \xLeftrightarrow[  ]{(\ref{substitution variable rearranged})}
\frac{\mu \sigma_v^2 - \left( y\sqrt{\sigma_v^2 + \sigma_u^2}-\mu \right) \sigma_u^2}{\sqrt{\sigma_v^2 + \sigma_u^2} \sigma_v \sigma_u} 
\Leftrightarrow \underbrace{\frac{\mu \sqrt{\sigma_v^2 + \sigma_u^2} }{\sigma_v \sigma_u}}_\text{a} + \underbrace{\left(- \frac{ \sigma_u }{\sigma_v } \right)}_\text{b} y  \\
\end{align*}

Substituting of $a,b$ and  $\varphi(\kappa)$ into the integral of Equation \ref{tnpdf} then yields \\ 
Lemma \ref{lemmatncdf}:
\[
\int_{-\infty}^{\kappa} f_\epsilon(t) dt=\frac{1}{\Phi (\frac{\mu}{\sigma_u})} \int_{-\infty}^{\varphi(\kappa)}    \phi \left( y \right) \Phi \left( a+by\right)  dy \\
\]
\end{proof}

\subsubsection{Proof of Theorem \ref{theoremtncdf}}
\begin{proof}
Theorem \ref{theoremtncdf} follows by applying Lemma \ref{lemmatncdf} and Equation 10,010.3 by \cite{owen1956tables}: 
\begin{equation} \label{Owen}
\begin{split}
\int \phi (y) \Phi (a+by) dy =& T\left(y, \frac{a}{y \sqrt{1+b^2}} \right) + T \left(\frac{a}{\sqrt{1+b^2}},\frac{y \sqrt{1+b^2}}{a} \right) \\
&-T \left( y,\frac{a+by}{y} \right)-T \left( \frac{a}{\sqrt{1+b^2}}, \frac{ab+y(1+b^2)}{a} \right) \\
&+ \Phi (y) \Phi \left( \frac{a}{\sqrt{1+b^2}} \right)
\end{split}
\end{equation}
in order to solve the obtained integral. An alternative representation is achieved by utilizing the following identity from \cite{owen1956tables}:
\[
T(h,g)=\frac{1}{2}\Phi(h)+\frac{1}{2}\Phi(gh)-\Phi(h)\Phi(gh)-T(gh,\frac{1}{g}) -\frac{1}{2} \mathbb{1}_{(-\infty,0)}(g) \text{ with } g\neq 0
\]
to rewrite the first term of Equation \ref{Owen} to:
\begin{align*}
T\left(y, \frac{a}{y \sqrt{1+b^2}} \right) &=  \frac{1}{2}\Phi(y)+\frac{1}{2}\Phi(\frac{a}{\sqrt{1+b^2}})-\Phi(y) \Phi(\frac{a}{\sqrt{1+b^2}}) \\
&-T(\frac{a}{\sqrt{1+b^2}},\frac{y \sqrt{1+b^2}}{a})-\frac{1}{2} \mathbb{1}_{(-\infty,0)}(\frac{a}{y \sqrt{1+b^2}})
\end{align*}

Equation \ref{Owen} can therefore be rewritten as:
\begin{align*}
=& \frac{1}{2}\Phi(y)+\frac{1}{2}\Phi(\frac{a}{\sqrt{1+b^2}}) -\frac{1}{2} \mathbb{1}_{(-\infty,0)}(\frac{a}{y \sqrt{1+b^2}})\\
&-T \left( y,\frac{a+by}{y} \right)-T \left( \frac{a}{\sqrt{1+b^2}}, \frac{ab+y(1+b^2)}{a} \right) \\
\end{align*}
Thus resulting in a compact representation of the integral of Equation \ref{tnpdf}, which from here on is referred to as \textit{Owen's T function CDF}:
\begin{align*} \label{Owen CDF}
F_\epsilon(\kappa)=&\frac{1}{\Phi (\frac{\mu}{\sigma_u})} \Bigg[ \frac{1}{2}\Phi(y)+\frac{1}{2}\Phi(\frac{a}{\sqrt{1+b^2}}) -\frac{1}{2} \mathbb{1}_{(-\infty,0)}(\frac{a}{y \sqrt{1+b^2}})\\
&-T \left( y,\frac{a+by}{y} \right)-T \left( \frac{a}{\sqrt{1+b^2}}, \frac{ab+y(1+b^2)}{a} \right) \Bigg]_{-\infty}^{\varphi(\kappa)}\\
=&\frac{1}{\Phi (\frac{\mu}{\sigma_u})} \Bigg[ \frac{1}{2}\Phi(\varphi(\kappa))+\frac{1}{2}\Phi(\frac{a}{\sqrt{1+b^2}}) -\frac{1}{2} \mathbb{1}_{(-\infty,0)}(\frac{a}{\varphi(\kappa) \sqrt{1+b^2}})\\
&-T \left( \varphi(\kappa),\frac{a+b\varphi(\kappa)}{\varphi(\kappa)} \right)-T \left( \frac{a}{\sqrt{1+b^2}}, \frac{ab+\varphi(\kappa)(1+b^2)}{a} \right) \Bigg] \\
\end{align*}
\end{proof}

Here it becomes clear that if as $\mu$ tends towards $0$, $a$ does the same, leading to a singularity. For the case $\mu=0$ the truncated normal distribution becomes the half-normal distribution, for which there is a closed form by \cite{amsler2019evaluating}. For the sake of completeness it is provided below:
\begin{align*}
F_\epsilon(\kappa ) &= 2 T(\frac{\kappa}{\sqrt{\sigma^2_{v}+\sigma^2_{u}}},\frac{\sigma_{u}}{\sigma_{v}})+\Phi(\frac{\kappa}{\sqrt{\sigma^2_{v}+\sigma^2_{u}}} )\\
\end{align*}
For $\lim_{t \to -\mu} y= 0$ the function exhibtis a singularity. 
\begin{align*}
\lim_{y \to 0} F_\epsilon(y)=&\frac{1}{\Phi (\frac{\mu}{\sigma_u})} \Bigg[ \frac{1}{4}+\frac{1}{2}\Phi(\frac{a}{\sqrt{1+b^2}})-\frac{1}{2} \mathbb{1}_{(-\infty,0)}(a)\\
&-\frac{1-\Phi(0)}{2}-T \left( \frac{a}{\sqrt{1+b^2}}, b \right) \Bigg] \\
\end{align*}

\subsubsection{Limiting Behavior}
The following equations, which can be found in \cite{owen1956tables} can be utilized to find the limits of the integral:
\begin{align*}
&\lim_{x \to -\infty} \Phi(x)=0 &\lim_{g \to \infty} T(h,g)=\frac{1-\Phi(|h|)}{2}  \\
&T \left( -h, g \right) = T \left( h, g \right) &T \left( h, -g \right)= -T \left( h, g \right)\\
&\lim_{h \to \infty} T( h, g) = 0  &T \left( 0, g\right)= \frac{arctan(g)}{2\pi}\\
\end{align*}

The functional value of the cdf as $\kappa$ tends towards $-\infty$ is:
\begin{align*}
\lim_{\kappa \to -\infty} F_\epsilon(\kappa)=& \lim_{\kappa \to -\infty} \frac{1}{\Phi (\frac{\mu}{\sigma_u})} \Bigg[ \frac{1}{2}\Phi(\kappa)+\frac{1}{2}\Phi(\frac{a}{\sqrt{1+b^2}}) -\frac{1}{2} \mathbb{1}_{(-\infty,0)}(\frac{a}{\kappa \sqrt{1+b^2}})\\
&-T \left( \kappa,\frac{a+b(\kappa)}{\kappa} \right)-T \left( \frac{a}{\sqrt{1+b^2}}, \frac{ab+(\kappa)(1+b^2)}{a} \right) \Bigg] \\
=& \lim_{\kappa \to -\infty}\frac{1}{\Phi (\frac{\mu}{\sigma_u})} \Bigg[ \frac{1}{2}\Phi(\frac{a}{\sqrt{1+b^2}}) -\frac{1}{2} \mathbb{1}_{(-\infty,0)}(\frac{a}{\kappa})\\
&-T \left( \kappa,b \right) - T \left( \frac{a}{\sqrt{1+b^2}}, \frac{\kappa}{a} \right) \Bigg] \\
=& \lim_{\kappa \to -\infty} \frac{1}{\Phi (\frac{\mu}{\sigma_u})} \Bigg[\frac{1}{2}\Phi(\frac{a}{\sqrt{1+b^2}}) -\frac{1}{2} \mathbb{1}_{(-\infty,0)}(\frac{a}{\kappa}) + sgn(a) \left( \frac{1}{2}-\frac{1}{2}\Phi(|\frac{a}{\sqrt{1+b^2}}|) \right) \Bigg] \\
\end{align*}

In the case of  $a<0$:
\begin{align*}
&\frac{1}{2} \Phi (\frac{a}{\sqrt(1 + b^2)}) +\left( \frac{1}{2} - \frac{1}{2} \Phi ( |\frac{a}{\sqrt{1 + b^2}} |)\right) = \frac{1}{2} \Phi (\frac{a}{\sqrt(1 + b^2)}) +\left( \frac{1}{2} - \frac{1}{2} (1 - \Phi ( \frac{a}{\sqrt{1 + b^2}} )\right) =0
\end{align*}

If $a>0$
\[
\frac{1}{2}\Phi(\frac{a}{\sqrt{1+b^2}})-\frac{1}{2} + \frac{1}{2}-\frac{1}{2}\Phi(\frac{a}{\sqrt{1+b^2}}) = 0
\]
\newline

The functional value of the cdf as $\kappa$ tends torwards $\infty$ is:
\begin{align*}
\lim_{\kappa \to \infty} F_\epsilon(\kappa)=& \lim_{\kappa \to \infty} \frac{1}{\Phi (\frac{\mu}{\sigma_u})} \Bigg[ \frac{1}{2}\Phi(\kappa)+\frac{1}{2}\Phi(\frac{a}{\sqrt{1+b^2}}) -\frac{1}{2} \mathbb{1}_{(-\infty,0)}(\frac{a}{\kappa \sqrt{1+b^2}})\\
&-T \left( \kappa,\frac{a+b(\kappa)}{\kappa} \right)-T \left( \frac{a}{\sqrt{1+b^2}}, \frac{ab+(\kappa)(1+b^2)}{a} \right) \Bigg] \\
=& \lim_{\kappa \to \infty} \frac{1}{\Phi (\frac{\mu}{\sigma_u})} \Bigg[ \frac{1}{2} -\frac{1}{2} \mathbb{1}_{(-\infty,0)}(\frac{a}{\kappa}) + \frac{1}{2} \Phi(\frac{a}{\sqrt{1+b^2}})  - sgn(a) \left( \frac{1}{2}-\frac{1}{2}\Phi(|\frac{a}{\sqrt{1+b^2}}|) \right) \Bigg] \\
\end{align*}

If $a>0$
\[
\frac{\Phi \left( \frac{a}{\sqrt{1+b^2}} \right)}{\Phi (\frac{\mu}{\sigma_u})}=1\\
\]

If $a<0$
\begin{align*}
\frac{1}{2} -\frac{1}{2} +\frac{1}{2}\Phi \left( \frac{a}{\sqrt{1+b^2}} \right) - (-1) \left(\frac{1}{2}-\frac{1}{2} (1 - \Phi\left( \frac{a}{\sqrt{1+b^2}}) \right) \right) =1\\
\end{align*}

For $\lim_{\sigma_u \to 0}$, $u$ becomes a degenerate random variable, i.e. deterministically assumes value $0$. Thus $\epsilon \sim N(-\mu_u,\sigma_v^2)$. \newline
Further, if $\lim_{\sigma_v \to 0}$, $v$ becomes a degenerate random variable taking value $-\mu$. Thus $\epsilon \sim TN(-\mu_u,\sigma_u^2,0,\infty)$.

\newpage 

\subsection{Representation using the Bivariate Normal Distribution}
\begin{theorem} \label{theoremtncdfbvn}
Let $u \sim TN(\mu,\sigma_u)$ and $v \sim N(0, \sigma_v)$ be independent, then it holds that the cdf of $\epsilon=v-u$ can be represented as:
\begin{align*}
F_\epsilon(\kappa) dt &=\frac{1}{\Phi (\frac{\mu}{\sigma_u})} BvN \left( \frac{\frac{\mu \sqrt{\sigma_v^2 + \sigma_u^2} }{\sigma_v \sigma_u}}{\sqrt{1+\left(- \frac{ \sigma_u }{\sigma_v } \right)^2}},  \frac{t+\mu}{\sqrt{\sigma_v^2 + \sigma_u^2}} , \rho=\frac{-\left(- \frac{ \sigma_u }{\sigma_v } \right)}{\sqrt{1+\left(- \frac{ \sigma_u }{\sigma_v } \right)^2}} \right)
\end{align*} 
where $BvN(\cdot)$ is the cdf of a bivariate normal distribution with correlation parameter $\rho \in [-1,1]$.
\end{theorem}

\subsubsection{Proof of Theorem \ref{theoremtncdfbvn}}
A similar approach to the proof of Theorem \ref{theoremtncdf} can be used to proof \\
Theorem \ref{theoremtncdfbvn}.
\begin{proof}
The Theorem \ref{theoremtncdfbvn} follows by applying  equation $10,010.3$ by  \cite{owen1980table}:
\begin{align*} 
\int_{-\infty}^{\varphi(\kappa)} \phi (y) \Phi (a+b y) dy &= \frac{1}{2 \pi \sqrt{1-\rho^2}} \int_{-\infty}^{\varphi(\kappa)} \int_{-\infty}^{\frac{a}{\sqrt{1+b^2}}} \exp \left[ -\frac{r^2-2 \rho rs+s^2}{2(1-\rho^2)} \right] dr ds\\
&=BvN \left( \frac{a}{\sqrt{1+b^2}}, \varphi(\kappa), \rho=\frac{-b}{\sqrt{1+b^2}} \right)
\end{align*}
to the result of Lemma \ref{lemmatncdf}, thus constructing a representation in terms of the cdf of the bivariate normal distribution.
Utilizing the introduced substitions from Lemma \ref{lemmatncdf}, the equation simplifies to 
\[
F_\epsilon(\kappa)=\frac{1}{\Phi (\frac{\mu}{\sigma_u})} \left[ BvN \left( \frac{a}{\sqrt{1+b^2}}, \varphi(\kappa), \rho=\frac{-b}{\sqrt{1+b^2}} \right) \right]
\]
which is referred to as \textit{BvN CDF}.
\end{proof}

\newpage

\section{Exponential Inefficiency Model} \label{Exponential Model}
In the following section, two representations of $F_\epsilon(\cdot)$ with $u \sim Exp(\lambda)$ are introduced. Further, the proofs are provided. Additionally, information on the limiting behavior is given.

\subsection{Representation using $exp$ function}

\begin{theorem} \label{theoremexpcdf}
Let $u \sim Exp(\lambda)$ and $v \sim N(0, \sigma_v)$ be independent, then it holds that the cdf of $\epsilon=v-u$ can be represented as:
\begin{align*}
F_\epsilon(\kappa) dt = 1 + \exp \{ -\frac{a^2 }{2}  \} \left[ \exp \{ a \varphi(\kappa) \} \Phi(y) -  \exp\{ \frac{a^2}{2} \} \Phi(\varphi(\kappa) -a) \right]  \\
\end{align*}
where $a=-\lambda \sigma_v$ and $\varphi(\kappa)= - \left( \frac{\kappa+\lambda \sigma_v^2}{\sigma_v} \right) $
\end{theorem}
Theorem \ref{theoremexpcdf} is a direct consequence of the following Lemma:
\begin{lemma}  \label{lemmaexpcdf}
Let $u \sim Exp(\lambda)$ and $v \sim N(0, \sigma_v)$ be independent, then it holds that the cdf of $\epsilon=v-u$ can be represented as:
\begin{align*}
\int_{-\infty}^{\kappa} f_\epsilon(t) dt &=-a \exp \{ -\frac{a^2 }{2} \} \int_{\varphi(\kappa)}^{\infty}  \exp \{  a y \} \Phi \left( y \right) dy 
\end{align*} 
where $a=-\lambda \sigma_v$ and $\varphi(\kappa)= - \left( \frac{\kappa+\lambda \sigma_v^2}{\sigma_v} \right) $.
\end{lemma}
Initially Lemma \ref{lemmaexpcdf} is proven and then it is shown how Theorem \ref{theoremexpcdf} follows. \newline

\subsubsection{Proof of Lemma \ref{lemmaexpcdf}}
\begin{proof}
Utilising standard algebra, the Equation \ref{exppdf} can be rearranged as follows:
\begin{align*}
&\lambda  \int_{-\infty}^{\kappa}  \exp \{ \lambda t \}  \exp \{ \frac{\sigma_v^2 \lambda^2 }{2}  \} \exp \{ -\lambda^2  \sigma_v^2 \} \exp \{ \lambda^2  \sigma_v^2 \}\Phi \left( -\frac{t+\lambda \sigma_v^2}{\sigma_v} \right) dt\\
& \Leftrightarrow \lambda   \exp \{ \frac{\sigma_v^2 \lambda^2 }{2}  - \lambda^2  \sigma_v^2 \} \int_{-\infty}^{\kappa}  \exp \{ \lambda t +\lambda^2 \sigma_v^2 \} \Phi \left( -\frac{t+\lambda \sigma_v^2}{\sigma_v} \right) dt\\
& \Leftrightarrow \lambda   \exp \{ -\frac{\sigma_v^2 \lambda^2 }{2} \} \int_{-\infty}^{\kappa}  \exp \bigg\{  \underbrace{-\lambda \sigma_v}_\text{a} \left( -\frac{ t +\lambda \sigma_v^2}{\sigma_v} \right) \bigg\} \Phi \left(  \underbrace{1}_\text{b} \left( -\frac{t+\lambda \sigma_v^2}{\sigma_v} \right) \right) dt\\
\end{align*}

Given the cdf as constructed through the integral of Equation \ref{exppdf}, the expression may be simplified by substition:
\[
y= \varphi(t):= - \left( \frac{t+\lambda \sigma_v^2}{\sigma_v} \right) \vspace{0.5cm},
\]
which can be rearranged as: 
\[
t= -y\sigma_v-\lambda \sigma_v^2 \\
\]

The derivative of $y$ w.r.t. $t$ is:
\[
\frac{d y}{d t}= -\frac{1}{\sigma_v} \leftrightarrow dt= -\sigma_v dy \vspace{0.5cm}.
\]
Appropriately transforming the limits of the integral results in:
\begin{align*}
&\varphi(\kappa) = - \left( \frac{\kappa +\lambda \sigma_v^2}{\sigma_v} \right)& \lim_{\kappa \to -\infty} \varphi(\kappa) =\infty
\end{align*}

Substituting $a,b, dt$ and $\varphi(\kappa)$ in the integral of Equation \ref{exppdf} then yields Lemma \ref{lemmaexpcdf}:
\begin{align*}
\int_{-\infty}^{\kappa} f_\epsilon(t) dt &= -a \exp \{ -\frac{a^2 }{2} \} \int_{\varphi(\kappa)}^{\infty}  \exp \{  a y \} \Phi \left( y \right) dy \\
\end{align*}
\end{proof}

\subsubsection{Proof of Theorem \ref{theoremexpcdf}}
\begin{proof}
Theorem \ref{theoremexpcdf} follows immediately by applying Lemma \ref{lemmaexpcdf} and Equation $101,000$ by \cite{owen1980table}:
\begin{align*}
\int \exp\{ a y\} \Phi(y)=\frac{1}{a} \exp \{ ay \} \Phi(b y) - \frac{1}{a} \exp\{ \frac{a^2}{2b^2} \} \Phi(by -\frac{a}{b}) \text{ with } ab \neq 0 \hspace{0.5cm}.
\end{align*}
Thus resulting in  a compact representation of the integral in Equation \ref{exppdf}:
\begin{align*}
F_\epsilon(\kappa)&=-a    \exp \{ -\frac{a^2 }{2}  \} \left[ \frac{1}{a} \exp \{ ay \} \Phi(by) - \frac{1}{a} \exp\{ \frac{a^2}{2b^2} \} \Phi(by -\frac{a}{b}) \right]^{\infty}_{\varphi(\kappa)} \\
&= 1 + \exp \{ -\frac{a^2 }{2}  \} \left[ \exp \{ a \varphi(\kappa) \} \Phi(y) -  \exp\{ \frac{a^2}{2} \} \Phi(\varphi(\kappa) -a) \right]  \\
\end{align*}
\end{proof}



\subsubsection{Limiting Behaviour}
The limits can be simplified with 
\begin{align*}
&\lim_{\kappa \to -\infty} \exp(\kappa) = 0 &\lim_{\kappa \to -\infty} \Phi(\kappa)= 0 \\
&\lim_{\kappa \to -\infty} \exp(-\kappa)\Phi(\kappa)= 0  &\lim_{\kappa \to \infty} \exp(\kappa)\Phi(-\kappa)= 0  \\
\end{align*}
.

The functional value of the cdf as $\kappa$ tends torwards $-\infty$ is:
\begin{align*}
\lim_{\kappa \to -\infty} F_\epsilon(\kappa) &= \lim_{\kappa \to -\infty} \left( 1 + \exp \{ -\frac{a^2 }{2}  \} \left[ \exp \{ a \varphi(\kappa) \} \Phi(\varphi(\kappa)) -  \exp\{ \frac{a^2}{2} \} \Phi(\varphi(\kappa) -a) \right] \right) \\
&=  1 + \exp \{ -\frac{a^2 }{2}  \} \left[ -  \exp\{ \frac{a^2}{2} \} \right] \\
&= 1 + (-1) = 0 
\end{align*} \newline
The functional value of the cdf as $\kappa$ tends torwards $\infty$ is:
\begin{align*}
\lim_{\kappa \to \infty} F_\epsilon(\kappa) &= \lim_{\kappa \to \infty} \left( 1 + \exp \{ -\frac{a^2 }{2}  \} \left[ \exp \{ a \varphi(\kappa) \} \Phi(\varphi(\kappa)) -  \exp\{ \frac{a^2}{2} \} \Phi(\varphi(\kappa) -a) \right] \right) \\
&=  1 + 0 =1 \\
\end{align*} \newline

For $\lim_{\lambda \to 0}$, $u$ becomes a degenerate random variable, i.e. deterministically assumes value $0$. Thus $\epsilon \sim N(0,\sigma_v^2)$.

\subsection{Representation using the Exponentially Modified Gaussian Distribution}
\begin{theorem} \label{theoremexpcdfemg}
Let $u \sim Exp(\lambda)$ and $v \sim N(0, \sigma_v)$ be independent, then it holds that the cdf of $\epsilon=v-u$ can be represented as:
\begin{align*}
F_\epsilon(\kappa) dt =1-F_{\epsilon^*}(-\kappa) \\
\end{align*}
where $F_{\epsilon^*}(\cdot)$ is cdf of the Exponentially Modified Gaussian Distribution with parameters $\mu=0,\sigma>0$ and $\lambda >0$.
\end{theorem}
The Exponentially Modified Gaussian (EMG)  distributed random variable $\epsilon^*$  is the sum of an independent normal and an exponential random variables. Thus
\begin{align*}
\epsilon^*&=v+u \vspace{0.5cm}. \\
\end{align*}
The cdf of the EMG with the mean of Gaussian component being $0$ is:
\[
F_{\epsilon^*}(\kappa)=\Phi\left(\lambda \kappa \right)-\exp \left( -\lambda \kappa+ (\lambda\sigma_v )^2/2 + \log(\Phi(\lambda \kappa,  (\lambda\sigma_v )^2,  \lambda\sigma_v )\right)
\]
introduced by \cite{gladney196911}.

\subsubsection{Proof of Theorem \ref{theoremexpcdfemg}}
\begin{proof}\label{prooftheorem4}
Since the random variable $v$ is symmetric around zero $-v$ and $v$ follow the same distribution, i.e. $-v\sim N(0,\sigma_v^2)$ and $v\sim N(0,\sigma_v^2)$. Consequently 
\[ \epsilon^*=-\epsilon=-v-u \vspace{0.5cm}. \]
Thus  
\[
F_{\epsilon}(\kappa)=P(\epsilon \leq \kappa)=P(-\epsilon \geq -\kappa)=1-P(-\epsilon \leq -\kappa)=1-F_{-\epsilon}(-\kappa)=1-F_{\epsilon^*}(-\kappa) \vspace{0.5cm}. 
\] 
This distribution is referred to as \textit{EmG CDF}.
\end{proof} 

\section{Simulation} \label{Simulation}
Validation of  the results is done by comparing the function values of the analytical cdfs to the function values of the empirical cdf. Construction of the  empirical cdf $F^*_\epsilon (\cdot)$ was done by drawing random numbers were drawn from $v$ and $u$. The cdfs were evaluated at the empirical quantiles $Q^*(p)$ with $p \in \{0.01, 0.05, 0.1, 0.25, 0.5, 0.75, 0.9, 0.95, 0.99 \}$ for any permutations of parameter values:
\begin{align*}
&\mu \in \{-8,-4,-2,-1,1,2,4,8 \} &\sigma_u \in \{0.25,0.5,1,2,4\} \\
&\lambda \in \{0.25,0.5,1,2,4, 8\} &\sigma_v \in \{0.25,0.5,1,2,4\}  
\end{align*}
The accuracy of the implementation is defined as:
\[
F_\epsilon (Q^*(p))-F^*_\epsilon (Q^*(p))=F_\epsilon (Q^*(p))-p \vspace{0.5cm}.
\] For each simulation scenario $10.000.000$ observations were generated, as in \cite{amsler2019evaluating}. Further, the accuracy and computation time of the of the representations of the cdfs are compared to numerical integration. Here, the double exponential integration method was chosen as it is considered a fairly good numerical integration method by \cite{Weisstein}.

\subsection{Simulation Results for the Truncated Normal Inefficiency Model}
\begin{figure}[H] 
\label{owenbvncomparison}
  \centering
\includegraphics[width=0.75\textwidth]{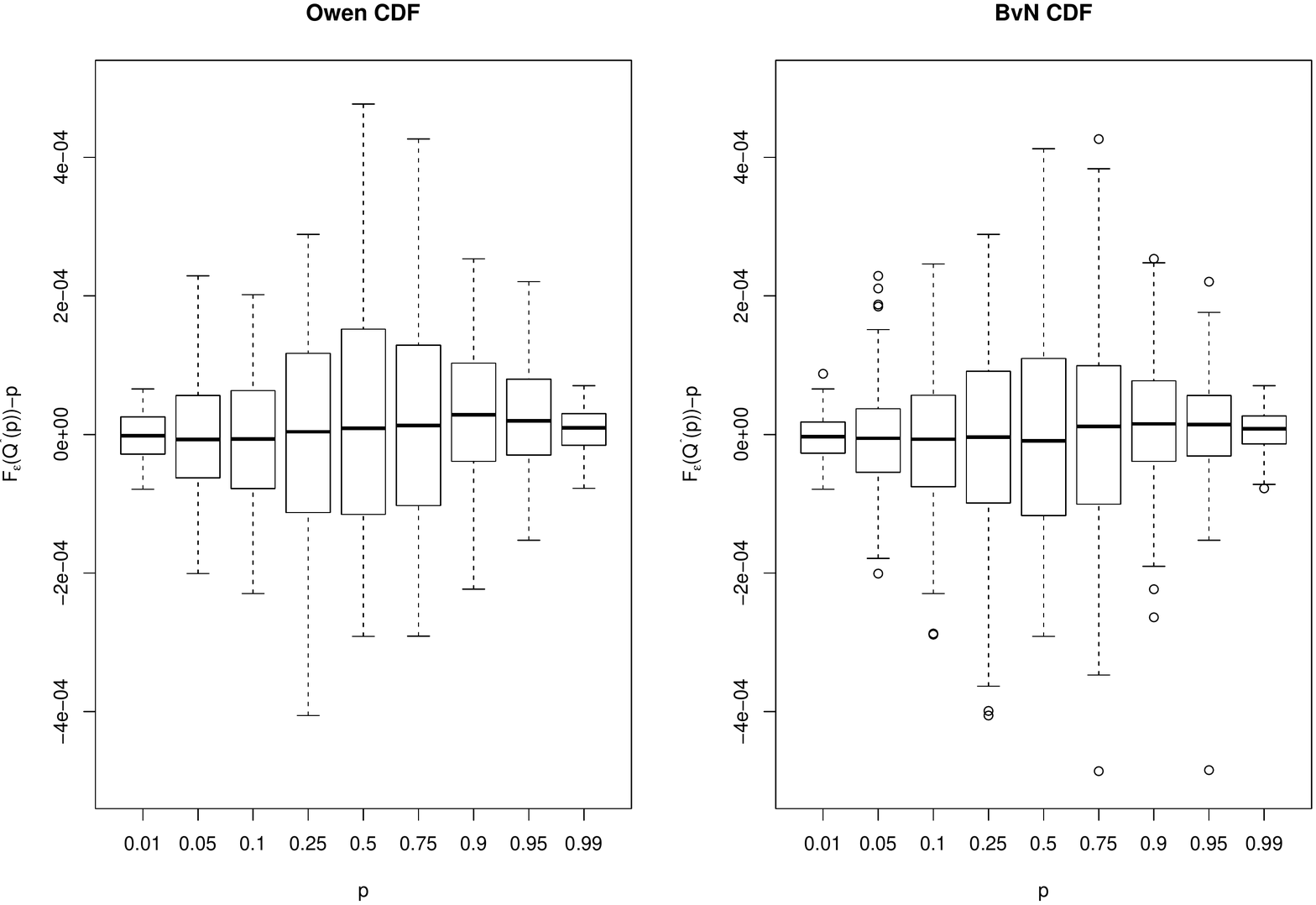} 
  \caption{Out of a total of 1800 evaluation points 269 outliers of the \textit{Owen CDF} and 12 outliers of \textit{BvN CDF} are not displayed in Figure \ref{owenbvncomparison}. The observed loss of accuracy for values close to $-\mu$ is due to described singularity. The \textit{BvN CDF} yields more accurate results.}
\end{figure}

For the truncated normal inefficiency model the derived formulas are identical in theory, but the accuracy of the numerical implementation does depend on both the implementation of the \\
Owen's T function and the bivariate normal cdf \footnote{The statistical software \textit{R} ($3.6.2$) was utilized. The truncated normal distributed random numbers were generated using the implementation of an an accept-reject sampler in the package \textit{truncnorm} ($1.0.8$). The Owen's T function of the \textit{pracma}($2.2.9$) and cdf of the bivariate normal distribution of the \textit{pbivnorm} ($0.6.0$) package were used.}. The figure \ref{owenbvncomparison} summarises the differences over all parameter combinations for different values of $p$ for the representations \textit{Owen CDF} and \textit{BvN CDF}. 
The implementation of the \textit{BvN CDF} is more accurate in this simulation. Thus, in the further analysis the \textit{Owen CDF} is neglected.  

In Table \ref{numbvncomparison} the relative accuracy and relative computation time of the numerical integration implementation relative to \textit{BvN CDF} is presented \footnote{For the numerical integration the \textit{pracma} package's function \textit{quadinf} was used. To measure the time the package \textit{microbenchmark}($1.4.7$) was utilized}.

\begin{table}[H]
  \centering
\caption{Comparison Numerical Integration to \textit{BvN CDF}}
\label{numbvncomparison}      
\begin{tabular}{lll}
\hline\noalign{\smallskip}
&  rel. accuracy & rel. time \\
\noalign{\smallskip}\hline\noalign{\smallskip}
 Min&  0.1156 & 42.16  \\
 1st Quartile & 1.000000  &  64.21 \\
 Median & 1.000000 &  73.63   \\
 Mean  \footnotemark &  0.99970 &  71.61  \\
 3rd Quartile &  1.000000 & 76.73  \\
 Max & 6.4027  & 101.87 \\
\noalign{\smallskip}\hline
\end{tabular}
\end{table}

\footnotetext{ If both the minimum and maximum of accuracy measure value are removed, the mean accuracy becomes $1.0034$} 
The results show that the \textit{BvN CDF} is faster in terms of computation time.

\subsection{Simulation Results for the Exponential Inefficiency Model}
For the exponential inefficiency model, both representations seem identical in terms of accuracy. The simulation results are \footnote{For the cdf of the EmG distribution the package \textit{emg} ($1.0.8$) was used}:
\begin{figure}[H] 
  \centering
\includegraphics[width=0.75\textwidth]{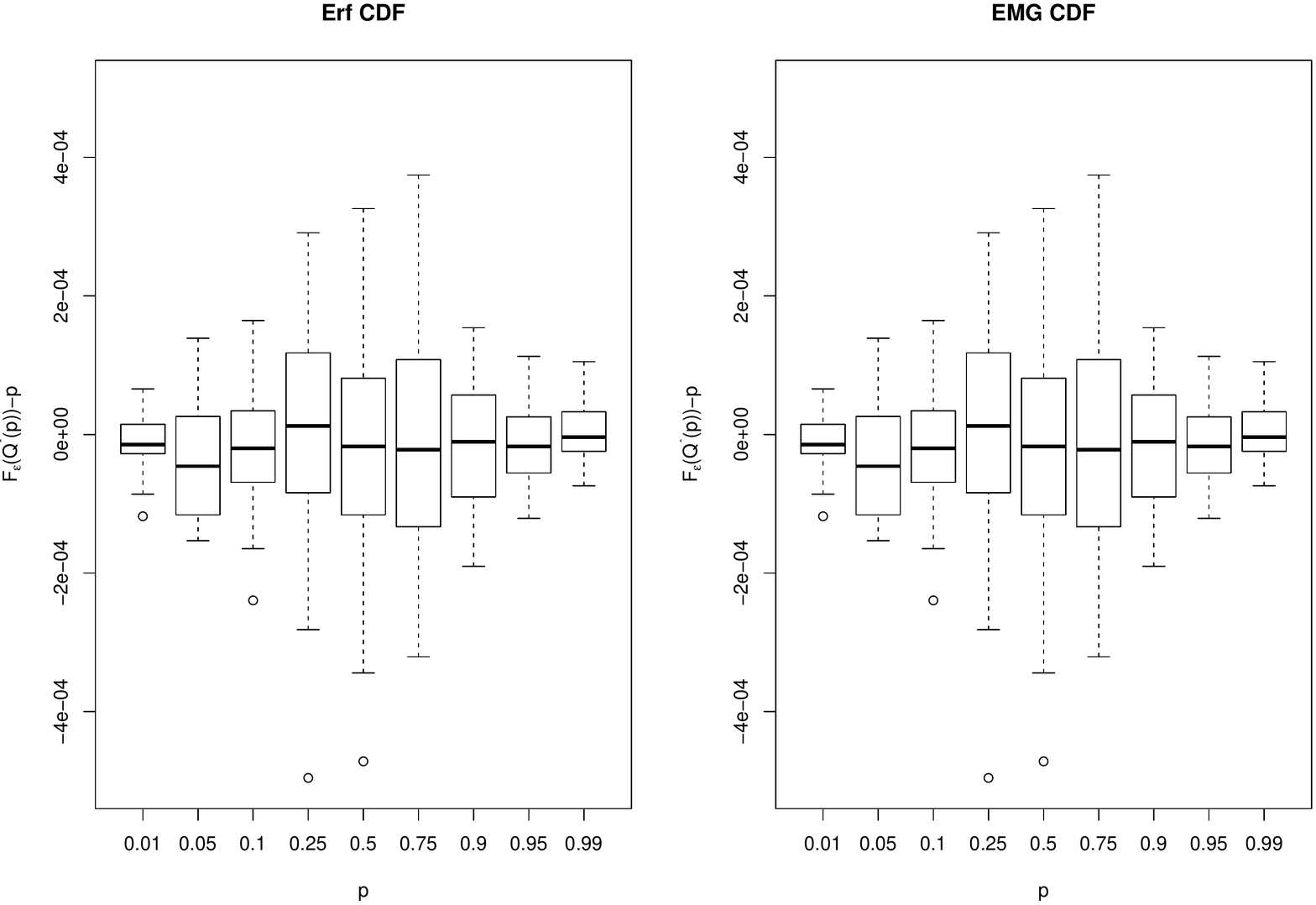} 
  \caption{The accuracy of both implementations seem not to differ.}
\end{figure} 
The implementation of the \textit{EmG CDF} is slightly faster. Thus the further analysis will focus on this representation.

\begin{table}[H]
  \centering
\caption{Comparison Numerical Integration to \textit{EmG CDF}}
\label{numemgcomparison}      
\begin{tabular}{lll}
\hline\noalign{\smallskip}
&  rel accuracy & rel time \\
\noalign{\smallskip}\hline\noalign{\smallskip}
 Min& 1.000000 & 689.4  \\
 1st Quartile & 1.000000  &  1008.0 \\
 Median & 1.000000 &  1190.3   \\
 Mean & 1.000000 &  1182.6   \\
 3rd Quartile &  1.000000 & 1365.4  \\
 Max & 1.000000  & 1923.0 \\
\noalign{\smallskip}\hline
\end{tabular}
\end{table}
The results in Table \ref{numemgcomparison} show that the \textit{EmG CDF} is equally accurate and faster in terms of computation time compared to the numerical integration. The accuracy is close to the numerical integration. 

A more detailed table of the simulation results for the \textit{BvN CDF} and \textit{EmG CDF} is presented in Section \ref{simtable}.

\newpage

\section{Conclusions}
The contribution of this paper are the analytical integrals of the cdf of the composed error term for the case of the inefficiency term following a truncated normal or exponential distribution. For the truncated normal inefficiency model the cdf can be written as the \textit{Owen CDF} and \textit{BvN CDF}, which are analytically the same but the numerical implementation of the latter is more accurate. In the exponential inefficiency model, the cdf is written as the \textit{Erf CDF} and \textit{EmG CDF}, which yield similiar results, in terms of accuracy with the second being faster to compute. The analytical representation of the cdfs allow for accurate and fast evaluation. 
\newpage

\begin{acknowledgements}
The authors would like to thank Alexander Ritz for his insights and helpful comments.
His mathematical support played an integral part in the derivation of the presented work.

The authors received financial support from the German Research Foundation (DFG) within the research project KN 922/9-1
\end{acknowledgements}

%
\section*{Conflict of interest}
The authors declare that they have no conflict of interest.

\newpage

\bibliographystyle{spbasic}      
\bibliography{SFA_CET_Paper}   

\begin{thebibliography}{15}
\providecommand{\natexlab}[1]{#1}
\providecommand{\url}[1]{{#1}}
\providecommand{\urlprefix}{URL }
\expandafter\ifx\csname urlstyle\endcsname\relax
  \providecommand{\doi}[1]{DOI~\discretionary{}{}{}#1}\else
  \providecommand{\doi}{DOI~\discretionary{}{}{}\begingroup
  \urlstyle{rm}\Url}\fi
\providecommand{\eprint}[2][]{\url{#2}}

\bibitem[{Amsler et~al.(2014)Amsler, Prokhorov, and Schmidt}]{amsler2014using}
Amsler C, Prokhorov A, Schmidt P (2014) Using copulas to model time dependence
  in stochastic frontier models. Econometric Reviews 33(5-6):497--522

\bibitem[{Amsler et~al.(2019)Amsler, Schmidt, and Tsay}]{amsler2019evaluating}
Amsler C, Schmidt P, Tsay WJ (2019) Evaluating the cdf of the distribution of
  the stochastic frontier composed error. Journal of Productivity Analysis
  52(1-3):29--35

\bibitem[{Genius et~al.(2012)Genius, Stefanou, and
  Tzouvelekas}]{genius2012measuring}
Genius M, Stefanou SE, Tzouvelekas V (2012) Measuring productivity growth under
  factor non-substitution: An application to us steam-electric power generation
  utilities. European journal of operational research 220(3):844--852

\bibitem[{Gladney(1969)}]{gladney196911}
Gladney H (1969) 11. f. dowden, jd swalen. Anal Chem 41:883

\bibitem[{Kumbhakar et~al.(2015)Kumbhakar, Wang, and
  Horncastle}]{kumbhakar2015practitioner}
Kumbhakar SC, Wang HJ, Horncastle AP (2015) A practitioner's guide to
  stochastic frontier analysis using Stata. Cambridge University Press

\bibitem[{Lai and Huang(2013)}]{lai2013maximum}
Lai Hp, Huang CJ (2013) Maximum likelihood estimation of seemingly unrelated
  stochastic frontier regressions. Journal of Productivity Analysis 40(1):1--14

\bibitem[{Meesters(2014)}]{meesters2014note}
Meesters A (2014) A note on the assumed distributions in stochastic frontier
  models. Journal of Productivity Analysis 42(2):171--173

\bibitem[{Meeusen and van Den~Broeck(1977)}]{meeusen1977efficiency}
Meeusen W, van Den~Broeck J (1977) Efficiency estimation from cobb-douglas
  production functions with composed error. International economic review pp
  435--444

\bibitem[{Owen(1956)}]{owen1956tables}
Owen DB (1956) Tables for computing bivariate normal probabilities. The Annals
  of Mathematical Statistics 27(4):1075--1090

\bibitem[{Owen(1980)}]{owen1980table}
Owen DB (1980) A table of normal integrals: A table. Communications in
  Statistics-Simulation and Computation 9(4):389--419

\bibitem[{Sriboonchitta et~al.(2017)Sriboonchitta, Liu, Wiboonpongse, and
  Denoeux}]{sriboonchitta2017double}
Sriboonchitta S, Liu J, Wiboonpongse A, Denoeux T (2017) A double-copula
  stochastic frontier model with dependent error components and correction for
  sample selection. International Journal of Approximate Reasoning 80:174--184

\bibitem[{Stevenson(1980)}]{stevenson1980likelihood}
Stevenson RE (1980) Likelihood functions for generalized stochastic frontier
  estimation. Journal of econometrics 13(1):57--66

\bibitem[{Tran and Tsionas(2015)}]{tran2015endogeneity}
Tran KC, Tsionas EG (2015) Endogeneity in stochastic frontier models: Copula
  approach without external instruments. Economics Letters 133:85--88

\bibitem[{Tsay et~al.(2013)Tsay, Huang, Fu, and Ho}]{tsay2013simple}
Tsay WJ, Huang CJ, Fu TT, Ho IL (2013) A simple closed-form approximation for
  the cumulative distribution function of the composite error of stochastic
  frontier models. Journal of Productivity Analysis 39(3):259--269

\bibitem[{Weisstein(1999)}]{Weisstein}
Weisstein EW (1999) Double exponential integration.
  \urlprefix\url{https://mathworld.wolfram.com/DoubleExponentialIntegration.html},
  visited on 2012-03-20

\end{thebibliography}

\section{Appendix}
\subsection{Simulation Result Tables} \label{simtable}
\begin{table}[H]
\centering
\caption{$|F_\epsilon(Q^*(p))-p|$ for $\mu \in \{-8,-4\},  \sigma_u \in \{0.25,0.5,1,2,4\}$ and $\sigma_v \in \{0.25,0.5,1,2,4\}$}
\resizebox{\textwidth}{!}{
\begin{tabular}{rrrrrrrrrrrr}
\hline\noalign{\smallskip}
$\mu$ & $\sigma_u$ & $\sigma_v$ & p=0.01 & p=0.05 & p=0.1 & p=0.25 & p=0.5 & p=0.75 & p=0.9 & p=0.95 & p=0.99 \\ 
\noalign{\smallskip}\hline\noalign{\smallskip}
-8 & 0.2 & 0.25 & 0.000005 & 0.000147 & 0.000287 & 0.000104 & 0.001174 & 0.000486 & 0.006924 & 0.006000 & 0.010000 \\ 
  -8 & 0.2 & 0.50 & 0.000012 & 0.000138 & 0.000246 & 0.000263 & 0.000186 & 0.000074 & 0.000264 & 0.000484 & 0.000684 \\ 
  -8 & 0.2 & 1.00 & 0.000011 & 0.000010 & 0.000289 & 0.000757 & 0.000604 & 0.002656 & 0.003326 & 0.007290 & 0.002980 \\ 
  -8 & 0.2 & 2.00 & 0.000035 & 0.000057 & 0.000189 & 0.000286 & 0.000030 & 0.000001 & 0.000049 & 0.000015 & 0.000003 \\ 
  -8 & 0.2 & 4.00 & 0.000021 & 0.000016 & 0.000004 & 0.000084 & 0.000088 & 0.000027 & 0.000100 & 0.000081 & 0.000002 \\ 
  -8 & 0.5 & 0.25 & 0.000032 & 0.000049 & 0.000089 & 0.000091 & 0.000043 & 0.000231 & 0.000028 & 0.000011 & 0.000010 \\ 
  -8 & 0.5 & 0.50 & 0.000008 & 0.000101 & 0.000014 & 0.000120 & 0.000169 & 0.000018 & 0.000110 & 0.000053 & 0.000063 \\ 
  -8 & 0.5 & 1.00 & 0.000020 & 0.000007 & 0.000031 & 0.000153 & 0.000065 & 0.000174 & 0.000248 & 0.000038 & 0.000030 \\ 
  -8 & 0.5 & 2.00 & 0.000037 & 0.000187 & 0.000178 & 0.000066 & 0.000030 & 0.000081 & 0.000109 & 0.000087 & 0.000007 \\ 
  -8 & 0.5 & 4.00 & 0.000026 & 0.000007 & 0.000002 & 0.000053 & 0.000218 & 0.000159 & 0.000039 & 0.000114 & 0.000023 \\ 
  -8 & 1.0 & 0.25 & 0.000009 & 0.000001 & 0.000110 & 0.000143 & 0.000041 & 0.000017 & 0.000082 & 0.000071 & 0.000012 \\ 
  -8 & 1.0 & 0.50 & 0.000049 & 0.000080 & 0.000001 & 0.000120 & 0.000141 & 0.000020 & 0.000050 & 0.000035 & 0.000057 \\ 
  -8 & 1.0 & 1.00 & 0.000017 & 0.000016 & 0.000005 & 0.000005 & 0.000100 & 0.000026 & 0.000046 & 0.000087 & 0.000015 \\ 
  -8 & 1.0 & 2.00 & 0.000065 & 0.000094 & 0.000041 & 0.000108 & 0.000191 & 0.000216 & 0.000136 & 0.000073 & 0.000031 \\ 
  -8 & 1.0 & 4.00 & 0.000088 & 0.000211 & 0.000215 & 0.000238 & 0.000145 & 0.000111 & 0.000009 & 0.000066 & 0.000030 \\ 
  -8 & 2.0 & 0.25 & 0.000015 & 0.000061 & 0.000091 & 0.000120 & 0.000036 & 0.000202 & 0.000108 & 0.000116 & 0.000042 \\ 
  -8 & 2.0 & 0.50 & 0.000024 & 0.000018 & 0.000122 & 0.000137 & 0.000278 & 0.000170 & 0.000008 & 0.000082 & 0.000006 \\ 
  -8 & 2.0 & 1.00 & 0.000004 & 0.000089 & 0.000082 & 0.000063 & 0.000066 & 0.000078 & 0.000026 & 0.000048 & 0.000005 \\ 
  -8 & 2.0 & 2.00 & 0.000040 & 0.000120 & 0.000088 & 0.000100 & 0.000039 & 0.000244 & 0.000006 & 0.000051 & 0.000031 \\ 
  -8 & 2.0 & 4.00 & 0.000024 & 0.000014 & 0.000014 & 0.000078 & 0.000156 & 0.000075 & 0.000109 & 0.000055 & 0.000062 \\ 
  -8 & 4.0 & 0.25 & 0.000028 & 0.000081 & 0.000129 & 0.000035 & 0.000147 & 0.000053 & 0.000023 & 0.000051 & 0.000031 \\ 
  -8 & 4.0 & 0.50 & 0.000031 & 0.000136 & 0.000099 & 0.000024 & 0.000062 & 0.000185 & 0.000165 & 0.000130 & 0.000047 \\ 
  -8 & 4.0 & 1.00 & 0.000025 & 0.000046 & 0.000104 & 0.000363 & 0.000291 & 0.000026 & 0.000103 & 0.000003 & 0.000007 \\ 
  -8 & 4.0 & 2.00 & 0.000017 & 0.000002 & 0.000114 & 0.000051 & 0.000107 & 0.000208 & 0.000010 & 0.000084 & 0.000018 \\ 
  -8 & 4.0 & 4.00 & 0.000001 & 0.000052 & 0.000040 & 0.000010 & 0.000071 & 0.000148 & 0.000039 & 0.000042 & 0.000068 \\ 
  -4 & 0.2 & 0.25 & 0.000013 & 0.000179 & 0.000210 & 0.000011 & 0.000118 & 0.000119 & 0.000084 & 0.000020 & 0.000061 \\ 
  -4 & 0.2 & 0.50 & 0.000041 & 0.000090 & 0.000188 & 0.000165 & 0.000201 & 0.000155 & 0.000091 & 0.000085 & 0.000036 \\ 
  -4 & 0.2 & 1.00 & 0.000033 & 0.000067 & 0.000106 & 0.000040 & 0.000132 & 0.000073 & 0.000005 & 0.000020 & 0.000019 \\ 
  -4 & 0.2 & 2.00 & 0.000008 & 0.000004 & 0.000069 & 0.000127 & 0.000078 & 0.000145 & 0.000006 & 0.000060 & 0.000014 \\ 
  -4 & 0.2 & 4.00 & 0.000033 & 0.000049 & 0.000094 & 0.000076 & 0.000285 & 0.000168 & 0.000112 & 0.000115 & 0.000005 \\ 
  -4 & 0.5 & 0.25 & 0.000027 & 0.000025 & 0.000079 & 0.000061 & 0.000232 & 0.000087 & 0.000049 & 0.000023 & 0.000015 \\ 
  -4 & 0.5 & 0.50 & 0.000007 & 0.000026 & 0.000022 & 0.000091 & 0.000179 & 0.000022 & 0.000000 & 0.000022 & 0.000025 \\ 
  -4 & 0.5 & 1.00 & 0.000034 & 0.000142 & 0.000117 & 0.000009 & 0.000028 & 0.000122 & 0.000073 & 0.000084 & 0.000041 \\ 
  -4 & 0.5 & 2.00 & 0.000034 & 0.000027 & 0.000060 & 0.000067 & 0.000117 & 0.000122 & 0.000085 & 0.000148 & 0.000004 \\ 
  -4 & 0.5 & 4.00 & 0.000023 & 0.000024 & 0.000112 & 0.000203 & 0.000093 & 0.000293 & 0.000162 & 0.000022 & 0.000015 \\ 
  -4 & 1.0 & 0.25 & 0.000026 & 0.000068 & 0.000037 & 0.000035 & 0.000012 & 0.000080 & 0.000072 & 0.000025 & 0.000002 \\ 
  -4 & 1.0 & 0.50 & 0.000029 & 0.000037 & 0.000121 & 0.000022 & 0.000162 & 0.000193 & 0.000083 & 0.000070 & 0.000023 \\ 
  -4 & 1.0 & 1.00 & 0.000047 & 0.000089 & 0.000106 & 0.000110 & 0.000191 & 0.000253 & 0.000081 & 0.000009 & 0.000049 \\ 
  -4 & 1.0 & 2.00 & 0.000033 & 0.000018 & 0.000014 & 0.000064 & 0.000117 & 0.000101 & 0.000026 & 0.000041 & 0.000033 \\ 
  -4 & 1.0 & 4.00 & 0.000022 & 0.000090 & 0.000056 & 0.000008 & 0.000108 & 0.000074 & 0.000062 & 0.000064 & 0.000026 \\ 
  -4 & 2.0 & 0.25 & 0.000010 & 0.000014 & 0.000104 & 0.000095 & 0.000134 & 0.000057 & 0.000062 & 0.000023 & 0.000005 \\ 
  -4 & 2.0 & 0.50 & 0.000041 & 0.000105 & 0.000035 & 0.000099 & 0.000007 & 0.000041 & 0.000015 & 0.000057 & 0.000005 \\ 
  -4 & 2.0 & 1.00 & 0.000030 & 0.000085 & 0.000202 & 0.000210 & 0.000256 & 0.000384 & 0.000085 & 0.000041 & 0.000006 \\ 
  -4 & 2.0 & 2.00 & 0.000027 & 0.000085 & 0.000003 & 0.000045 & 0.000175 & 0.000258 & 0.000117 & 0.000022 & 0.000005 \\ 
  -4 & 2.0 & 4.00 & 0.000001 & 0.000006 & 0.000063 & 0.000068 & 0.000000 & 0.000045 & 0.000111 & 0.000034 & 0.000018 \\ 
  -4 & 4.0 & 0.25 & 0.000079 & 0.000101 & 0.000105 & 0.000082 & 0.000061 & 0.000274 & 0.000180 & 0.000118 & 0.000025 \\ 
  -4 & 4.0 & 0.50 & 0.000058 & 0.000018 & 0.000019 & 0.000066 & 0.000097 & 0.000213 & 0.000046 & 0.000014 & 0.000016 \\ 
  -4 & 4.0 & 1.00 & 0.000058 & 0.000012 & 0.000046 & 0.000189 & 0.000135 & 0.000232 & 0.000229 & 0.000067 & 0.000004 \\ 
  -4 & 4.0 & 2.00 & 0.000016 & 0.000069 & 0.000060 & 0.000034 & 0.000047 & 0.000013 & 0.000030 & 0.000026 & 0.000016 \\ 
  -4 & 4.0 & 4.00 & 0.000018 & 0.000185 & 0.000160 & 0.000212 & 0.000377 & 0.000119 & 0.000033 & 0.000029 & 0.000022 \\ 
\noalign{\smallskip}\hline
\end{tabular}}
\end{table}

\begin{table}[H]
\centering
\caption{$|F_\epsilon(Q^*(p))-p|$ for $\mu \in \{-2,-1\},  \sigma_u \in \{0.25,0.5,1,2,4\}$ and $\sigma_v \in \{0.25,0.5,1,2,4\}$}
\resizebox{\textwidth}{!}{
\begin{tabular}{rrrrrrrrrrrr}
\hline\noalign{\smallskip}
$\mu$ & $\sigma_u$ & $\sigma_v$ & p=0.01 & p=0.05 & p=0.1 & p=0.25 & p=0.5 & p=0.75 & p=0.9 & p=0.95 & p=0.99 \\ 
\noalign{\smallskip}\hline\noalign{\smallskip}
 -2 & 0.2 & 0.25 & 0.000027 & 0.000072 & 0.000101 & 0.000243 & 0.000195 & 0.000120 & 0.000003 & 0.000007 & 0.000013 \\ 
  -2 & 0.2 & 0.50 & 0.000020 & 0.000001 & 0.000019 & 0.000063 & 0.000018 & 0.000165 & 0.000007 & 0.000065 & 0.000025 \\ 
  -2 & 0.2 & 1.00 & 0.000010 & 0.000013 & 0.000009 & 0.000039 & 0.000160 & 0.000347 & 0.000161 & 0.000001 & 0.000002 \\ 
  -2 & 0.2 & 2.00 & 0.000001 & 0.000004 & 0.000099 & 0.000071 & 0.000029 & 0.000160 & 0.000100 & 0.000034 & 0.000040 \\ 
  -2 & 0.2 & 4.00 & 0.000008 & 0.000094 & 0.000150 & 0.000107 & 0.000156 & 0.000001 & 0.000052 & 0.000002 & 0.000025 \\ 
  -2 & 0.5 & 0.25 & 0.000009 & 0.000094 & 0.000078 & 0.000013 & 0.000076 & 0.000188 & 0.000047 & 0.000010 & 0.000018 \\ 
  -2 & 0.5 & 0.50 & 0.000027 & 0.000001 & 0.000112 & 0.000192 & 0.000251 & 0.000296 & 0.000181 & 0.000029 & 0.000014 \\ 
  -2 & 0.5 & 1.00 & 0.000048 & 0.000093 & 0.000060 & 0.000220 & 0.000114 & 0.000069 & 0.000004 & 0.000010 & 0.000006 \\ 
  -2 & 0.5 & 2.00 & 0.000025 & 0.000078 & 0.000139 & 0.000066 & 0.000156 & 0.000138 & 0.000048 & 0.000062 & 0.000008 \\ 
  -2 & 0.5 & 4.00 & 0.000000 & 0.000007 & 0.000069 & 0.000026 & 0.000052 & 0.000152 & 0.000035 & 0.000064 & 0.000028 \\ 
  -2 & 1.0 & 0.25 & 0.000019 & 0.000201 & 0.000152 & 0.000230 & 0.000052 & 0.000185 & 0.000181 & 0.000170 & 0.000003 \\ 
  -2 & 1.0 & 0.50 & 0.000027 & 0.000021 & 0.000016 & 0.000025 & 0.000041 & 0.000245 & 0.000198 & 0.000043 & 0.000015 \\ 
  -2 & 1.0 & 1.00 & 0.000004 & 0.000096 & 0.000108 & 0.000118 & 0.000057 & 0.000021 & 0.000043 & 0.000060 & 0.000024 \\ 
  -2 & 1.0 & 2.00 & 0.000010 & 0.000055 & 0.000128 & 0.000142 & 0.000007 & 0.000119 & 0.000040 & 0.000002 & 0.000032 \\ 
  -2 & 1.0 & 4.00 & 0.000056 & 0.000013 & 0.000041 & 0.000061 & 0.000049 & 0.000111 & 0.000024 & 0.000042 & 0.000034 \\ 
  -2 & 2.0 & 0.25 & 0.000060 & 0.000085 & 0.000124 & 0.000069 & 0.000141 & 0.000047 & 0.000033 & 0.000002 & 0.000003 \\ 
  -2 & 2.0 & 0.50 & 0.000062 & 0.000080 & 0.000022 & 0.000058 & 0.000049 & 0.000015 & 0.000029 & 0.000090 & 0.000039 \\ 
  -2 & 2.0 & 1.00 & 0.000028 & 0.000122 & 0.000055 & 0.000098 & 0.000200 & 0.000010 & 0.000017 & 0.000028 & 0.000045 \\ 
  -2 & 2.0 & 2.00 & 0.000027 & 0.000005 & 0.000026 & 0.000053 & 0.000031 & 0.000171 & 0.000054 & 0.000087 & 0.000036 \\ 
  -2 & 2.0 & 4.00 & 0.000012 & 0.000033 & 0.000036 & 0.000100 & 0.000042 & 0.000044 & 0.000009 & 0.000015 & 0.000025 \\ 
  -2 & 4.0 & 0.25 & 0.000031 & 0.000079 & 0.000009 & 0.000097 & 0.000189 & 0.000012 & 0.000060 & 0.000004 & 0.000003 \\ 
  -2 & 4.0 & 0.50 & 0.000011 & 0.000018 & 0.000082 & 0.000193 & 0.000062 & 0.000005 & 0.000006 & 0.000012 & 0.000001 \\ 
  -2 & 4.0 & 1.00 & 0.000002 & 0.000005 & 0.000017 & 0.000175 & 0.000108 & 0.000251 & 0.000040 & 0.000069 & 0.000011 \\ 
  -2 & 4.0 & 2.00 & 0.000009 & 0.000058 & 0.000010 & 0.000071 & 0.000103 & 0.000181 & 0.000023 & 0.000084 & 0.000024 \\ 
  -2 & 4.0 & 4.00 & 0.000029 & 0.000017 & 0.000027 & 0.000012 & 0.000080 & 0.000086 & 0.000048 & 0.000012 & 0.000025 \\ 
  -1 & 0.2 & 0.25 & 0.000039 & 0.000122 & 0.000108 & 0.000191 & 0.000342 & 0.000085 & 0.000003 & 0.000012 & 0.000013 \\ 
  -1 & 0.2 & 0.50 & 0.000015 & 0.000061 & 0.000058 & 0.000236 & 0.000168 & 0.000027 & 0.000005 & 0.000066 & 0.000012 \\ 
  -1 & 0.2 & 1.00 & 0.000027 & 0.000052 & 0.000090 & 0.000155 & 0.000099 & 0.000291 & 0.000090 & 0.000080 & 0.000007 \\ 
  -1 & 0.2 & 2.00 & 0.000051 & 0.000164 & 0.000152 & 0.000128 & 0.000010 & 0.000142 & 0.000000 & 0.000008 & 0.000014 \\ 
  -1 & 0.2 & 4.00 & 0.000036 & 0.000028 & 0.000064 & 0.000311 & 0.000184 & 0.000047 & 0.000081 & 0.000026 & 0.000048 \\ 
  -1 & 0.5 & 0.25 & 0.000001 & 0.000001 & 0.000067 & 0.000019 & 0.000128 & 0.000047 & 0.000020 & 0.000010 & 0.000009 \\ 
  -1 & 0.5 & 0.50 & 0.000013 & 0.000048 & 0.000042 & 0.000169 & 0.000145 & 0.000198 & 0.000190 & 0.000141 & 0.000028 \\ 
  -1 & 0.5 & 1.00 & 0.000029 & 0.000026 & 0.000083 & 0.000030 & 0.000063 & 0.000054 & 0.000034 & 0.000051 & 0.000001 \\ 
  -1 & 0.5 & 2.00 & 0.000016 & 0.000028 & 0.000004 & 0.000111 & 0.000133 & 0.000054 & 0.000113 & 0.000031 & 0.000008 \\ 
  -1 & 0.5 & 4.00 & 0.000005 & 0.000023 & 0.000010 & 0.000091 & 0.000181 & 0.000162 & 0.000223 & 0.000137 & 0.000039 \\ 
  -1 & 1.0 & 0.25 & 0.000011 & 0.000086 & 0.000000 & 0.000032 & 0.000016 & 0.000083 & 0.000223 & 0.000153 & 0.000064 \\ 
  -1 & 1.0 & 0.50 & 0.000024 & 0.000038 & 0.000064 & 0.000049 & 0.000056 & 0.000097 & 0.000040 & 0.000013 & 0.000003 \\ 
  -1 & 1.0 & 1.00 & 0.000004 & 0.000053 & 0.000112 & 0.000160 & 0.000289 & 0.000248 & 0.000078 & 0.000018 & 0.000014 \\ 
  -1 & 1.0 & 2.00 & 0.000028 & 0.000074 & 0.000116 & 0.000194 & 0.000196 & 0.000046 & 0.000035 & 0.000031 & 0.000024 \\ 
  -1 & 1.0 & 4.00 & 0.000031 & 0.000034 & 0.000150 & 0.000264 & 0.000128 & 0.000107 & 0.000131 & 0.000066 & 0.000042 \\ 
  -1 & 2.0 & 0.25 & 0.000022 & 0.000027 & 0.000094 & 0.000029 & 0.000139 & 0.000086 & 0.000030 & 0.000083 & 0.000028 \\ 
  -1 & 2.0 & 0.50 & 0.000022 & 0.000050 & 0.000069 & 0.000008 & 0.000002 & 0.000020 & 0.000036 & 0.000041 & 0.000019 \\ 
  -1 & 2.0 & 1.00 & 0.000002 & 0.000046 & 0.000061 & 0.000169 & 0.000155 & 0.000175 & 0.000040 & 0.000101 & 0.000030 \\ 
  -1 & 2.0 & 2.00 & 0.000025 & 0.000056 & 0.000019 & 0.000172 & 0.000129 & 0.000055 & 0.000032 & 0.000004 & 0.000025 \\ 
  -1 & 2.0 & 4.00 & 0.000028 & 0.000054 & 0.000082 & 0.000060 & 0.000114 & 0.000177 & 0.000029 & 0.000113 & 0.000027 \\ 
  -1 & 4.0 & 0.25 & 0.000021 & 0.000151 & 0.000047 & 0.000112 & 0.000095 & 0.000131 & 0.000032 & 0.000023 & 0.000020 \\ 
  -1 & 4.0 & 0.50 & 0.000028 & 0.000062 & 0.000023 & 0.000063 & 0.000243 & 0.000198 & 0.000038 & 0.000026 & 0.000035 \\ 
  -1 & 4.0 & 1.00 & 0.000006 & 0.000022 & 0.000054 & 0.000149 & 0.000176 & 0.000046 & 0.000088 & 0.000038 & 0.000004 \\ 
  -1 & 4.0 & 2.00 & 0.000037 & 0.000127 & 0.000102 & 0.000050 & 0.000021 & 0.000013 & 0.000044 & 0.000020 & 0.000012 \\ 
  -1 & 4.0 & 4.00 & 0.000038 & 0.000067 & 0.000009 & 0.000145 & 0.000144 & 0.000186 & 0.000013 & 0.000044 & 0.000023 \\ 
\noalign{\smallskip}\hline
\end{tabular}}
\end{table}

\begin{table}[H]
\centering
\caption{$|F_\epsilon(Q^*(p))-p|$ for $\mu \in \{1,2\},  \sigma_u \in \{0.25,0.5,1,2,4\}$ and $\sigma_v \in \{0.25,0.5,1,2,4\}$}
\resizebox{\textwidth}{!}{
\begin{tabular}{rrrrrrrrrrrr}
 \hline\noalign{\smallskip}
$\mu$ & $\sigma_u$ & $\sigma_v$ & p=0.01 & p=0.05 & p=0.1 & p=0.25 & p=0.5 & p=0.75 & p=0.9 & p=0.95 & p=0.99 \\ 
\noalign{\smallskip}\hline\noalign{\smallskip}
  1 & 0.2 & 0.25 & 0.000008 & 0.000013 & 0.000109 & 0.000024 & 0.000052 & 0.000219 & 0.000064 & 0.000072 & 0.000026 \\ 
  1 & 0.2 & 0.50 & 0.000004 & 0.000008 & 0.000131 & 0.000399 & 0.000125 & 0.000028 & 0.000052 & 0.000068 & 0.000015 \\ 
  1 & 0.2 & 1.00 & 0.000042 & 0.000073 & 0.000109 & 0.000048 & 0.000105 & 0.000032 & 0.000006 & 0.000015 & 0.000004 \\ 
  1 & 0.2 & 2.00 & 0.000022 & 0.000089 & 0.000082 & 0.000148 & 0.000197 & 0.000127 & 0.000044 & 0.000062 & 0.000012 \\ 
  1 & 0.2 & 4.00 & 0.000048 & 0.000097 & 0.000187 & 0.000259 & 0.000023 & 0.000134 & 0.000114 & 0.000034 & 0.000050 \\ 
  1 & 0.5 & 0.25 & 0.000029 & 0.000005 & 0.000032 & 0.000180 & 0.000056 & 0.000221 & 0.000071 & 0.000057 & 0.000041 \\ 
  1 & 0.5 & 0.50 & 0.000031 & 0.000025 & 0.000045 & 0.000033 & 0.000263 & 0.000175 & 0.000222 & 0.000161 & 0.000022 \\ 
  1 & 0.5 & 1.00 & 0.000032 & 0.000121 & 0.000061 & 0.000018 & 0.000022 & 0.000006 & 0.000035 & 0.000145 & 0.000018 \\ 
  1 & 0.5 & 2.00 & 0.000028 & 0.000027 & 0.000080 & 0.000069 & 0.000223 & 0.000030 & 0.000059 & 0.000066 & 0.000006 \\ 
  1 & 0.5 & 4.00 & 0.000032 & 0.000015 & 0.000074 & 0.000017 & 0.000014 & 0.000070 & 0.000103 & 0.000028 & 0.000016 \\ 
  1 & 1.0 & 0.25 & 0.000021 & 0.000021 & 0.000096 & 0.000036 & 0.000235 & 0.000144 & 0.000123 & 0.000048 & 0.000010 \\ 
  1 & 1.0 & 0.50 & 0.000015 & 0.000229 & 0.000128 & 0.000192 & 0.000147 & 0.000142 & 0.000115 & 0.000020 & 0.000012 \\ 
  1 & 1.0 & 1.00 & 0.000040 & 0.000076 & 0.000127 & 0.000289 & 0.000249 & 0.000161 & 0.000106 & 0.000103 & 0.000056 \\ 
  1 & 1.0 & 2.00 & 0.000002 & 0.000010 & 0.000053 & 0.000106 & 0.000154 & 0.000082 & 0.000031 & 0.000001 & 0.000010 \\ 
  1 & 1.0 & 4.00 & 0.000068 & 0.000112 & 0.000051 & 0.000171 & 0.000073 & 0.000039 & 0.000009 & 0.000047 & 0.000062 \\ 
  1 & 2.0 & 0.25 & 0.000075 & 0.000108 & 0.000230 & 0.000269 & 0.000195 & 0.000052 & 0.000023 & 0.000021 & 0.000078 \\ 
  1 & 2.0 & 0.50 & 0.000024 & 0.000010 & 0.000121 & 0.000227 & 0.000041 & 0.000053 & 0.000021 & 0.000008 & 0.000015 \\ 
  1 & 2.0 & 1.00 & 0.000015 & 0.000005 & 0.000005 & 0.000103 & 0.000313 & 0.000072 & 0.000016 & 0.000073 & 0.000005 \\ 
  1 & 2.0 & 2.00 & 0.000008 & 0.000007 & 0.000010 & 0.000222 & 0.000098 & 0.000146 & 0.000117 & 0.000003 & 0.000014 \\ 
  1 & 2.0 & 4.00 & 0.000008 & 0.000037 & 0.000072 & 0.000108 & 0.000021 & 0.000053 & 0.000046 & 0.000001 & 0.000008 \\ 
  1 & 4.0 & 0.25 & 0.000039 & 0.000023 & 0.000063 & 0.000028 & 0.000141 & 0.000149 & 0.000037 & 0.000021 & 0.000008 \\ 
  1 & 4.0 & 0.50 & 0.000007 & 0.000032 & 0.000026 & 0.000208 & 0.000191 & 0.000062 & 0.000064 & 0.000061 & 0.000041 \\ 
  1 & 4.0 & 1.00 & 0.000012 & 0.000001 & 0.000024 & 0.000011 & 0.000165 & 0.000110 & 0.000028 & 0.000041 & 0.000050 \\ 
  1 & 4.0 & 2.00 & 0.000003 & 0.000017 & 0.000147 & 0.000033 & 0.000112 & 0.000031 & 0.000018 & 0.000006 & 0.000031 \\ 
  1 & 4.0 & 4.00 & 0.000004 & 0.000009 & 0.000037 & 0.000112 & 0.000154 & 0.000019 & 0.000179 & 0.000176 & 0.000028 \\ 
  2 & 0.2 & 0.25 & 0.000031 & 0.000077 & 0.000039 & 0.000093 & 0.000156 & 0.000057 & 0.000004 & 0.000080 & 0.000030 \\ 
  2 & 0.2 & 0.50 & 0.000050 & 0.000026 & 0.000128 & 0.000182 & 0.000036 & 0.000017 & 0.000043 & 0.000018 & 0.000012 \\ 
  2 & 0.2 & 1.00 & 0.000022 & 0.000030 & 0.000013 & 0.000106 & 0.000089 & 0.000068 & 0.000106 & 0.000026 & 0.000010 \\ 
  2 & 0.2 & 2.00 & 0.000023 & 0.000056 & 0.000013 & 0.000025 & 0.000199 & 0.000069 & 0.000052 & 0.000050 & 0.000015 \\ 
  2 & 0.2 & 4.00 & 0.000025 & 0.000086 & 0.000197 & 0.000257 & 0.000412 & 0.000164 & 0.000069 & 0.000016 & 0.000037 \\ 
  2 & 0.5 & 0.25 & 0.000014 & 0.000057 & 0.000039 & 0.000180 & 0.000039 & 0.000120 & 0.000088 & 0.000068 & 0.000025 \\ 
  2 & 0.5 & 0.50 & 0.000010 & 0.000040 & 0.000073 & 0.000025 & 0.000137 & 0.000002 & 0.000179 & 0.000064 & 0.000033 \\ 
  2 & 0.5 & 1.00 & 0.000068 & 0.000119 & 0.000148 & 0.000049 & 0.000141 & 0.000133 & 0.000136 & 0.000100 & 0.000021 \\ 
  2 & 0.5 & 2.00 & 0.000010 & 0.000048 & 0.000065 & 0.000157 & 0.000164 & 0.000041 & 0.000004 & 0.000012 & 0.000027 \\ 
  2 & 0.5 & 4.00 & 0.000041 & 0.000082 & 0.000131 & 0.000214 & 0.000018 & 0.000093 & 0.000116 & 0.000069 & 0.000040 \\ 
  2 & 1.0 & 0.25 & 0.000047 & 0.000118 & 0.000059 & 0.000064 & 0.000149 & 0.000050 & 0.000033 & 0.000065 & 0.000003 \\ 
  2 & 1.0 & 0.50 & 0.000006 & 0.000070 & 0.000002 & 0.000118 & 0.000068 & 0.000080 & 0.000014 & 0.000052 & 0.000066 \\ 
  2 & 1.0 & 1.00 & 0.000017 & 0.000052 & 0.000064 & 0.000052 & 0.000040 & 0.000100 & 0.000027 & 0.000017 & 0.000003 \\ 
  2 & 1.0 & 2.00 & 0.000010 & 0.000057 & 0.000059 & 0.000121 & 0.000129 & 0.000111 & 0.000093 & 0.000008 & 0.000015 \\ 
  2 & 1.0 & 4.00 & 0.000029 & 0.000039 & 0.000070 & 0.000015 & 0.000007 & 0.000011 & 0.000046 & 0.000107 & 0.000031 \\ 
  2 & 2.0 & 0.25 & 0.000027 & 0.000058 & 0.000026 & 0.000061 & 0.000011 & 0.000060 & 0.000119 & 0.000129 & 0.000015 \\ 
  2 & 2.0 & 0.50 & 0.000009 & 0.000061 & 0.000048 & 0.000046 & 0.000043 & 0.000079 & 0.000043 & 0.000051 & 0.000058 \\ 
  2 & 2.0 & 1.00 & 0.000026 & 0.000046 & 0.000041 & 0.000040 & 0.000067 & 0.000061 & 0.000006 & 0.000048 & 0.000032 \\ 
  2 & 2.0 & 2.00 & 0.000008 & 0.000092 & 0.000063 & 0.000076 & 0.000007 & 0.000108 & 0.000119 & 0.000037 & 0.000014 \\ 
  2 & 2.0 & 4.00 & 0.000004 & 0.000059 & 0.000220 & 0.000318 & 0.000141 & 0.000103 & 0.000111 & 0.000111 & 0.000013 \\ 
  2 & 4.0 & 0.25 & 0.000003 & 0.000057 & 0.000061 & 0.000001 & 0.000100 & 0.000169 & 0.000097 & 0.000030 & 0.000003 \\ 
  2 & 4.0 & 0.50 & 0.000031 & 0.000007 & 0.000049 & 0.000003 & 0.000129 & 0.000080 & 0.000090 & 0.000098 & 0.000017 \\ 
  2 & 4.0 & 1.00 & 0.000008 & 0.000087 & 0.000150 & 0.000164 & 0.000000 & 0.000119 & 0.000020 & 0.000033 & 0.000036 \\ 
  2 & 4.0 & 2.00 & 0.000039 & 0.000045 & 0.000005 & 0.000292 & 0.000023 & 0.000049 & 0.000018 & 0.000014 & 0.000013 \\ 
  2 & 4.0 & 4.00 & 0.000010 & 0.000022 & 0.000068 & 0.000060 & 0.000088 & 0.000171 & 0.000108 & 0.000088 & 0.000008 \\ 
\noalign{\smallskip}\hline
\end{tabular}}
\end{table}

\begin{table}[H]
\centering
\caption{$|F_\epsilon(Q^*(p))-p|$ for $\mu \in \{4,8\},  \sigma_u \in \{0.25,0.5,1,2,4\}$ and $\sigma_v \in \{0.25,0.5,1,2,4\}$}
\resizebox{\textwidth}{!}{
\begin{tabular}{rrrrrrrrrrrr}
 \hline\noalign{\smallskip}
$\mu$ & $\sigma_u$ & $\sigma_v$ & p=0.01 & p=0.05 & p=0.1 & p=0.25 & p=0.5 & p=0.75 & p=0.9 & p=0.95 & p=0.99 \\ 
\noalign{\smallskip}\hline\noalign{\smallskip}
 4 & 0.2 & 0.25 & 0.000009 & 0.000023 & 0.000143 & 0.000033 & 0.000132 & 0.000080 & 0.000011 & 0.000030 & 0.000011 \\ 
  4 & 0.2 & 0.50 & 0.000029 & 0.000024 & 0.000039 & 0.000054 & 0.000191 & 0.000053 & 0.000076 & 0.000068 & 0.000024 \\ 
  4 & 0.2 & 1.00 & 0.000021 & 0.000025 & 0.000023 & 0.000106 & 0.000290 & 0.000170 & 0.000088 & 0.000084 & 0.000070 \\ 
  4 & 0.2 & 2.00 & 0.000039 & 0.000124 & 0.000196 & 0.000019 & 0.000181 & 0.000238 & 0.000129 & 0.000146 & 0.000018 \\ 
  4 & 0.2 & 4.00 & 0.000026 & 0.000027 & 0.000002 & 0.000200 & 0.000119 & 0.000101 & 0.000071 & 0.000041 & 0.000063 \\ 
  4 & 0.5 & 0.25 & 0.000014 & 0.000121 & 0.000018 & 0.000016 & 0.000180 & 0.000193 & 0.000122 & 0.000018 & 0.000033 \\ 
  4 & 0.5 & 0.50 & 0.000020 & 0.000107 & 0.000088 & 0.000146 & 0.000274 & 0.000086 & 0.000073 & 0.000131 & 0.000013 \\ 
  4 & 0.5 & 1.00 & 0.000036 & 0.000092 & 0.000171 & 0.000122 & 0.000189 & 0.000066 & 0.000083 & 0.000018 & 0.000038 \\ 
  4 & 0.5 & 2.00 & 0.000048 & 0.000052 & 0.000025 & 0.000225 & 0.000216 & 0.000248 & 0.000091 & 0.000115 & 0.000001 \\ 
  4 & 0.5 & 4.00 & 0.000010 & 0.000034 & 0.000096 & 0.000171 & 0.000056 & 0.000049 & 0.000030 & 0.000044 & 0.000022 \\ 
  4 & 1.0 & 0.25 & 0.000042 & 0.000064 & 0.000126 & 0.000029 & 0.000023 & 0.000013 & 0.000049 & 0.000101 & 0.000021 \\ 
  4 & 1.0 & 0.50 & 0.000044 & 0.000006 & 0.000115 & 0.000090 & 0.000039 & 0.000137 & 0.000050 & 0.000076 & 0.000062 \\ 
  4 & 1.0 & 1.00 & 0.000036 & 0.000035 & 0.000005 & 0.000036 & 0.000205 & 0.000048 & 0.000126 & 0.000081 & 0.000072 \\ 
  4 & 1.0 & 2.00 & 0.000005 & 0.000066 & 0.000060 & 0.000071 & 0.000013 & 0.000104 & 0.000136 & 0.000090 & 0.000044 \\ 
  4 & 1.0 & 4.00 & 0.000001 & 0.000079 & 0.000004 & 0.000297 & 0.000174 & 0.000137 & 0.000051 & 0.000035 & 0.000020 \\ 
  4 & 2.0 & 0.25 & 0.000003 & 0.000037 & 0.000006 & 0.000159 & 0.000094 & 0.000073 & 0.000084 & 0.000047 & 0.000032 \\ 
  4 & 2.0 & 0.50 & 0.000008 & 0.000015 & 0.000065 & 0.000020 & 0.000102 & 0.000098 & 0.000011 & 0.000084 & 0.000019 \\ 
  4 & 2.0 & 1.00 & 0.000015 & 0.000009 & 0.000011 & 0.000073 & 0.000129 & 0.000186 & 0.000009 & 0.000042 & 0.000029 \\ 
  4 & 2.0 & 2.00 & 0.000033 & 0.000082 & 0.000008 & 0.000118 & 0.000111 & 0.000144 & 0.000051 & 0.000052 & 0.000054 \\ 
  4 & 2.0 & 4.00 & 0.000004 & 0.000094 & 0.000157 & 0.000165 & 0.000288 & 0.000426 & 0.000190 & 0.000135 & 0.000018 \\ 
  4 & 4.0 & 0.25 & 0.000015 & 0.000060 & 0.000063 & 0.000101 & 0.000201 & 0.000244 & 0.000105 & 0.000076 & 0.000041 \\ 
  4 & 4.0 & 0.50 & 0.000036 & 0.000035 & 0.000039 & 0.000075 & 0.000043 & 0.000036 & 0.000115 & 0.000052 & 0.000021 \\ 
  4 & 4.0 & 1.00 & 0.000016 & 0.000005 & 0.000027 & 0.000160 & 0.000134 & 0.000161 & 0.000010 & 0.000044 & 0.000023 \\ 
  4 & 4.0 & 2.00 & 0.000046 & 0.000089 & 0.000151 & 0.000124 & 0.000055 & 0.000008 & 0.000090 & 0.000004 & 0.000065 \\ 
  4 & 4.0 & 4.00 & 0.000027 & 0.000033 & 0.000119 & 0.000088 & 0.000011 & 0.000058 & 0.000055 & 0.000079 & 0.000011 \\ 
  8 & 0.2 & 0.25 & 0.000058 & 0.000126 & 0.000074 & 0.000162 & 0.000257 & 0.000125 & 0.000128 & 0.000052 & 0.000012 \\ 
  8 & 0.2 & 0.50 & 0.000044 & 0.000030 & 0.000038 & 0.000206 & 0.000132 & 0.000181 & 0.000067 & 0.000048 & 0.000022 \\ 
  8 & 0.2 & 1.00 & 0.000005 & 0.000018 & 0.000039 & 0.000116 & 0.000275 & 0.000050 & 0.000057 & 0.000039 & 0.000043 \\ 
  8 & 0.2 & 2.00 & 0.000001 & 0.000038 & 0.000042 & 0.000144 & 0.000016 & 0.000194 & 0.000089 & 0.000220 & 0.000027 \\ 
  8 & 0.2 & 4.00 & 0.000066 & 0.000016 & 0.000000 & 0.000184 & 0.000696 & 0.000340 & 0.000080 & 0.000032 & 0.000026 \\ 
  8 & 0.5 & 0.25 & 0.000033 & 0.000041 & 0.000058 & 0.000122 & 0.000144 & 0.000130 & 0.000075 & 0.000012 & 0.000029 \\ 
  8 & 0.5 & 0.50 & 0.000022 & 0.000033 & 0.000002 & 0.000078 & 0.000228 & 0.000332 & 0.000253 & 0.000157 & 0.000003 \\ 
  8 & 0.5 & 1.00 & 0.000038 & 0.000019 & 0.000023 & 0.000005 & 0.000139 & 0.000057 & 0.000003 & 0.000076 & 0.000032 \\ 
  8 & 0.5 & 2.00 & 0.000039 & 0.000067 & 0.000185 & 0.000406 & 0.000125 & 0.000111 & 0.000103 & 0.000025 & 0.000004 \\ 
  8 & 0.5 & 4.00 & 0.000066 & 0.000076 & 0.000182 & 0.000143 & 0.000030 & 0.000045 & 0.000077 & 0.000019 & 0.000004 \\ 
  8 & 1.0 & 0.25 & 0.000035 & 0.000068 & 0.000087 & 0.000089 & 0.000102 & 0.000012 & 0.000108 & 0.000103 & 0.000009 \\ 
  8 & 1.0 & 0.50 & 0.000011 & 0.000008 & 0.000013 & 0.000063 & 0.000038 & 0.000060 & 0.000012 & 0.000023 & 0.000039 \\ 
  8 & 1.0 & 1.00 & 0.000006 & 0.000036 & 0.000078 & 0.000037 & 0.000041 & 0.000011 & 0.000048 & 0.000018 & 0.000042 \\ 
  8 & 1.0 & 2.00 & 0.000053 & 0.000014 & 0.000078 & 0.000171 & 0.000081 & 0.000193 & 0.000143 & 0.000140 & 0.000040 \\ 
  8 & 1.0 & 4.00 & 0.000029 & 0.000028 & 0.000011 & 0.000077 & 0.000023 & 0.000042 & 0.000162 & 0.000151 & 0.000010 \\ 
  8 & 2.0 & 0.25 & 0.000025 & 0.000041 & 0.000078 & 0.000028 & 0.000074 & 0.000142 & 0.000073 & 0.000038 & 0.000018 \\ 
  8 & 2.0 & 0.50 & 0.000018 & 0.000061 & 0.000022 & 0.000190 & 0.000063 & 0.000122 & 0.000122 & 0.000021 & 0.000002 \\ 
  8 & 2.0 & 1.00 & 0.000042 & 0.000115 & 0.000150 & 0.000225 & 0.000102 & 0.000095 & 0.000021 & 0.000087 & 0.000017 \\ 
  8 & 2.0 & 2.00 & 0.000021 & 0.000068 & 0.000066 & 0.000021 & 0.000019 & 0.000179 & 0.000116 & 0.000094 & 0.000043 \\ 
  8 & 2.0 & 4.00 & 0.000013 & 0.000015 & 0.000007 & 0.000013 & 0.000030 & 0.000035 & 0.000039 & 0.000025 & 0.000035 \\ 
  8 & 4.0 & 0.25 & 0.000009 & 0.000027 & 0.000037 & 0.000227 & 0.000019 & 0.000050 & 0.000049 & 0.000034 & 0.000019 \\ 
  8 & 4.0 & 0.50 & 0.000003 & 0.000003 & 0.000002 & 0.000113 & 0.000126 & 0.000052 & 0.000093 & 0.000098 & 0.000030 \\ 
  8 & 4.0 & 1.00 & 0.000007 & 0.000061 & 0.000093 & 0.000046 & 0.000063 & 0.000048 & 0.000049 & 0.000053 & 0.000011 \\ 
  8 & 4.0 & 2.00 & 0.000011 & 0.000011 & 0.000024 & 0.000030 & 0.000052 & 0.000125 & 0.000038 & 0.000013 & 0.000004 \\ 
  8 & 4.0 & 4.00 & 0.000044 & 0.000099 & 0.000115 & 0.000096 & 0.000003 & 0.000069 & 0.000051 & 0.000054 & 0.000008 \\
\noalign{\smallskip}\hline
\end{tabular}}
\end{table}

\begin{table}[H]
\centering
\caption{$|F_\epsilon(Q^*(p))-p|$ for $\lambda \in \{0.25,0.5,1,2,4,8\}$ and $\sigma_v \in \{0.25,0.5,1,2,4\}$}
\resizebox{1\textwidth}{!}{
\begin{tabular}{rrrrrrrrrrr}
\hline\noalign{\smallskip}
$\lambda$ & $\sigma_v$ & p=0.01 & p=0.05 & p=0.1 & p=0.25 & p=0.5 & p=0.75 & p=0.9 & p=0.95 & p=0.99 \\ 
\noalign{\smallskip}\hline\noalign{\smallskip}
0.25 & 0.25 & 0.000086 & 0.000150 & 0.000240 & 0.000262 & 0.000116 & 0.000138 & 0.000027 & 0.000052 & 0.000018 \\ 
  0.25 & 0.50 & 0.000007 & 0.000068 & 0.000095 & 0.000025 & 0.000029 & 0.000110 & 0.000086 & 0.000056 & 0.000036 \\ 
  0.25 & 1.00 & 0.000015 & 0.000018 & 0.000053 & 0.000043 & 0.000162 & 0.000127 & 0.000007 & 0.000012 & 0.000105 \\ 
  0.25 & 2.00 & 0.000037 & 0.000111 & 0.000058 & 0.000031 & 0.000034 & 0.000149 & 0.000069 & 0.000009 & 0.000026 \\ 
  0.25 & 4.00 & 0.000013 & 0.000048 & 0.000011 & 0.000053 & 0.000049 & 0.000037 & 0.000006 & 0.000074 & 0.000022 \\ 
  0.50 & 0.25 & 0.000068 & 0.000116 & 0.000004 & 0.000148 & 0.000081 & 0.000109 & 0.000116 & 0.000113 & 0.000007 \\ 
  0.50 & 0.50 & 0.000022 & 0.000126 & 0.000055 & 0.000046 & 0.000219 & 0.000108 & 0.000057 & 0.000100 & 0.000057 \\ 
  0.50 & 1.00 & 0.000028 & 0.000026 & 0.000069 & 0.000282 & 0.000197 & 0.000202 & 0.000182 & 0.000107 & 0.000057 \\ 
  0.50 & 2.00 & 0.000019 & 0.000140 & 0.000165 & 0.000271 & 0.000046 & 0.000088 & 0.000015 & 0.000064 & 0.000010 \\ 
  0.50 & 4.00 & 0.000012 & 0.000020 & 0.000094 & 0.000122 & 0.000092 & 0.000023 & 0.000185 & 0.000059 & 0.000007 \\ 
  1.00 & 0.25 & 0.000016 & 0.000121 & 0.000105 & 0.000100 & 0.000016 & 0.000194 & 0.000071 & 0.000082 & 0.000066 \\ 
  1.00 & 0.50 & 0.000066 & 0.000011 & 0.000112 & 0.000001 & 0.000229 & 0.000182 & 0.000154 & 0.000009 & 0.000050 \\ 
  1.00 & 1.00 & 0.000018 & 0.000051 & 0.000113 & 0.000291 & 0.000076 & 0.000321 & 0.000104 & 0.000046 & 0.000041 \\ 
  1.00 & 2.00 & 0.000003 & 0.000056 & 0.000005 & 0.000136 & 0.000326 & 0.000374 & 0.000147 & 0.000077 & 0.000033 \\ 
  1.00 & 4.00 & 0.000046 & 0.000020 & 0.000007 & 0.000025 & 0.000283 & 0.000092 & 0.000017 & 0.000025 & 0.000019 \\ 
  2.00 & 0.25 & 0.000008 & 0.000102 & 0.000027 & 0.000084 & 0.000096 & 0.000035 & 0.000018 & 0.000091 & 0.000023 \\ 
  2.00 & 0.50 & 0.000025 & 0.000017 & 0.000047 & 0.000102 & 0.000021 & 0.000040 & 0.000156 & 0.000056 & 0.000074 \\ 
  2.00 & 1.00 & 0.000007 & 0.000052 & 0.000085 & 0.000096 & 0.000009 & 0.000157 & 0.000003 & 0.000026 & 0.000041 \\ 
  2.00 & 2.00 & 0.000023 & 0.000036 & 0.000037 & 0.000060 & 0.000068 & 0.000030 & 0.000015 & 0.000012 & 0.000054 \\ 
  2.00 & 4.00 & 0.000022 & 0.000043 & 0.000035 & 0.000166 & 0.000202 & 0.000162 & 0.000191 & 0.000052 & 0.000038 \\ 
  4.00 & 0.25 & 0.000009 & 0.000113 & 0.000026 & 0.000129 & 0.000075 & 0.000024 & 0.000115 & 0.000121 & 0.000024 \\ 
  4.00 & 0.50 & 0.000040 & 0.000038 & 0.000034 & 0.000098 & 0.000160 & 0.000023 & 0.000057 & 0.000028 & 0.000019 \\ 
  4.00 & 1.00 & 0.000015 & 0.000067 & 0.000069 & 0.000070 & 0.000008 & 0.000133 & 0.000090 & 0.000070 & 0.000013 \\ 
  4.00 & 2.00 & 0.000024 & 0.000056 & 0.000127 & 0.000207 & 0.000171 & 0.000081 & 0.000095 & 0.000064 & 0.000067 \\ 
  4.00 & 4.00 & 0.000029 & 0.000132 & 0.000057 & 0.000083 & 0.000344 & 0.000202 & 0.000005 & 0.000044 & 0.000076 \\ 
  8.00 & 0.25 & 0.000065 & 0.000139 & 0.000164 & 0.000118 & 0.000053 & 0.000158 & 0.000106 & 0.000014 & 0.000001 \\ 
  8.00 & 0.50 & 0.000118 & 0.000154 & 0.000108 & 0.000084 & 0.000019 & 0.000142 & 0.000029 & 0.000023 & 0.000064 \\ 
  8.00 & 1.00 & 0.000052 & 0.000104 & 0.000080 & 0.000243 & 0.000171 & 0.000020 & 0.000028 & 0.000037 & 0.000022 \\ 
  8.00 & 2.00 & 0.000028 & 0.000126 & 0.000143 & 0.000496 & 0.000472 & 0.000203 & 0.000065 & 0.000019 & 0.000016 \\ 
  8.00 & 4.00 & 0.000019 & 0.000005 & 0.000001 & 0.000066 & 0.000317 & 0.000093 & 0.000018 & 0.000003 & 0.000005 \\ 
\noalign{\smallskip}\hline
\end{tabular}}
\end{table}

\end{document}